\documentclass[12pt]{article}
\def\int {\intop \limits}

\def\fnote#1{\footnote}
\begin{document}

\renewcommand \theequation{\thesection.\arabic{equation}}

\title{Polarization effects for pair creation 
by photon in oriented crystals at high energy}
\author{V. N. Baier
and V. M. Katkov\\
Budker Institute of Nuclear Physics\\ 630090 Novosibirsk, Russia}

\maketitle

\begin{abstract}
Pair creation by a photon in an oriented crystal can be considered
in a frame of the quasiclassical operator method, which includes
processes with polarized particles. Under some quite
generic assumptions the general expression is derived for the
probability of pair creation of longitudinally polarized electron
(positron) by circularly polarized photon in oriented crystal.
In particular cases $\vartheta_0 \ll V_0/m$ and 
$\vartheta_0 \gg V_0/m$~($\vartheta_0$ is 
the angle of incidence, angle between the
momentum of initial photon and axis (plane) of crystal, 
$V_0$ is the scale of a potential of
axis or a plane relative to which the angle $\vartheta_0$ is
defined) one has constant field approximation and the coherent 
pair production theory correspondingly. Side by side with
coherent process the probability of incoherent pair creation 
is calculated, which differs essentially from amorphous one.
At high energy the pair creation in oriented crystals is strongly 
enhanced comparing with amorphous medium.
In appendixes the integral 
polarization of positron is found in external field and for
coherent and incoherent mechanisms. 

\end{abstract}

\newpage

\section{Introduction}

The study of processes with participation of polarized electrons 
and photons permits to obtain the important physical information.
Because of this reason the experiments with use of polarized particles
are performed and are planning in many laboratories (CERN, Jefferson 
Nat Accl Fac, SLAC, BINP, etc). In this paper it is 
shown that oriented crystal is an unique tool for work with polarized
electrons and photons
 
The quasiclassical operator method developed by authors
\cite{BK0}-\cite{BK2} is adequate for consideration of the
electromagnetic processes at high energy. The probability of
polarized pair creation by a circularly polarized photon has a
form (see \cite{BKS}, p.73, Eq.(3.12); the method is given also in
\cite{BLP},\cite{BKF})
\begin{equation}
dw=\frac{e^2}{(2\pi)^2} \frac{d^3p}{\omega}
\int_{}^{}dt_2\int_{}^{}dt_1
R_{s\overline{s}}^{(\lambda)}(t_2)R_{s\overline{s}}^{(\lambda)\ast}(t_1)
\exp \left[\frac{i\varepsilon}{\varepsilon'}
\left(kx(t_2)-kx(t_1)\right) \right],
\label{1}
\end{equation}
where for polarized electrons and positrons one has within
the relativistic accuracy
\begin{equation}
R_{s\overline{s}}^{(\lambda)}=\frac{m}{\sqrt{\varepsilon
\varepsilon'}}u_s^{+}(p)\mathbf{e}^{\lambda}\mbox{\boldmath$\alpha$}
v_{\overline{s}}(p')
\label{2}
\end{equation}
here $k^{\nu}=(\omega, {\bf k})$ is the 4-momentum of the initial 
photon, $p^{\mu}=(\varepsilon, {\bf p})$ is the 4-momentum of the
created electron, $k^2=0$, $x^{\mu}(t)=(t, {\bf r}(t))$, $t$ is
the time, and ${\bf r}(t)$ is the created electron location on a
classical trajectory, $kx(t)=\omega t- {\bf kr}(t)$,
$\varepsilon'=\omega-\varepsilon,~
\mathbf{p}'=\mathbf{k}-\mathbf{p},~
\lambda=\pm 1$ is the circular polarization of photon,
$s(\overline{s})=\pm 1$ is the longitudinal(with respect to
direction of motion) polarization of electron (positron), we
employ units such that $\hbar=c=1$.

Passing on to two-component spinors $\varphi$ and choosing as the
quantization axis (the axis $z$) the direction of photon momentum
$\mathbf{k}$ we get
\begin{eqnarray}
&&
\mathbf{e}^{(\lambda)}=\frac{1}{\sqrt{2}}(\mathbf{e}_x+i\lambda\mathbf{e}_y),\quad
\mathbf{e}^{\lambda}\mbox{\boldmath$\alpha$}=\sqrt{2}
\left(\begin{array}{cc}
  0 & \hat{s}_{\lambda} \\
  \hat{s}_{\lambda} & 0
\end{array}\right),
\nonumber \\
&&
\sqrt{\frac{m}{\varepsilon}}u_s(p)=\sqrt{\frac{\varepsilon+m}{2\varepsilon}}
\left(\begin{array}{c}
  \varphi_s  \\
  \frac{\mbox{\boldmath$\sigma$}\textbf{p}} {\varepsilon+m}\varphi_s
\end{array}\right),
\nonumber \\
&& \sqrt{\frac{m}{\varepsilon'}}v_{\overline{s}}(p')
=\sqrt{\frac{\varepsilon'+m}{2\varepsilon'}}
\left(\begin{array}{c}
  \frac{\mbox{\boldmath$\sigma$}\textbf{p}'}{\varepsilon'+m}
  \varphi_{-\overline{s}}  \\
  \varphi_{-\overline{s}}
\end{array}\right).
\label{3}
\end{eqnarray}
Here $\hat{s}_{\pm}=
(\sigma_x \pm i\sigma_y)/2$ is the
raising (lowering) operator:
\begin{equation}
\sigma_z \varphi_s=s \varphi_s,\quad
\hat{s}_{\lambda}\varphi_s=\delta_{\lambda, -s}\varphi_{-s},\quad
\varphi_s^{+}\hat{s}_{\lambda}=\delta_{\lambda, s}\varphi_{-s}^{+}.
\label{4}
\end{equation}

Below we will consider the small angle approximation (the vector {\bf
p} is directed nearly along the vector {\bf k}, just this configuration
yields the main contribution into the process probability) and will retain
only the main terms in the decomposition over $1/\gamma~(\gamma=
\varepsilon/m$ is the Lorentz factor). Within
this accuracy one obtains
\begin{eqnarray}
&& \sqrt{\frac{m}{\varepsilon}}u_s(p)=\frac{1}{\sqrt{2}}
\left(\begin{array}{c}
  \varphi_s  \\
  \left[s\left(1-\frac{m}{\varepsilon}\right)
  +\mbox{\boldmath$\sigma$}\mbox{\boldmath$\vartheta$}\right]\varphi_s
\end{array}\right),
\nonumber \\
&& \sqrt{\frac{m}{\varepsilon'}}v_{\overline{s}}(p')
=\frac{1}{\sqrt{2}}
\left(\begin{array}{c}
  - \left[\overline{s}\left(1-\frac{m}{\varepsilon'}\right)
  +\frac{\varepsilon}{\varepsilon'}
  \mbox{\boldmath$\sigma$}\mbox{\boldmath$\vartheta$}\right]\varphi_{-\overline{s}}  \\
  \varphi_{-\overline{s}}
\end{array}\right),
\label{5}
\end{eqnarray}
where $\mbox{\boldmath$\vartheta$}=\textbf{v}-\textbf{n}=\textbf{v}_{\perp},
\textbf{v}=\textbf{v}(t)$ is the electron velocity, $\textbf{v}_{\perp}$ is the
component of electron velocity perpendicular to the vector 
$\textbf{n}=\textbf{k}/\omega$. Let us note that 
$\mbox{\boldmath$\vartheta'$}=\textbf{v}'-\textbf{n}=\textbf{p}'/\varepsilon'
-\textbf{n}=(\textbf{n}-\textbf{v})\varepsilon/\varepsilon',
~\vartheta'\gamma'=\vartheta\gamma$.

Using the relations
\begin{eqnarray}
&& \varphi_s^{+} \mbox{\boldmath$\sigma$}\mbox{\boldmath$\vartheta$}\varphi_s=0,
\quad \varphi_{-s}^{+} \mbox{\boldmath$\sigma$}\mbox{\boldmath$\vartheta$}\varphi_s=
\vartheta_x+is\vartheta_y \equiv \vartheta^{(s)},\quad
\varphi_s^{+} \hat{s}_{\lambda}\varphi_{-\overline{s}}
=\delta_{\lambda,s}\delta_{\lambda,\overline{s}},
\nonumber \\
&& \varphi_{s}^{+} \hat{s}_{\lambda}\mbox{\boldmath$\sigma$}
\mbox{\boldmath$\vartheta$}\varphi_{-\overline{s}}=\delta_{\lambda,s}
\varphi_{-s}^{+} \mbox{\boldmath$\sigma$}\mbox{\boldmath$\vartheta$}
\varphi_{-\overline{s}}
=\vartheta^{(\lambda)}\delta_{\lambda,s}\delta_{\lambda,-\overline{s}},
\nonumber \\
&& \varphi_{s}^{+} \mbox{\boldmath$\sigma$}
\mbox{\boldmath$\vartheta$}\hat{s}_{\lambda}\varphi_{-\overline{s}}=
\delta_{\lambda,\overline{s}}
\varphi_{s}^{+} \mbox{\boldmath$\sigma$}\mbox{\boldmath$\vartheta$}
\varphi_{\overline{s}}
=\vartheta^{(\lambda)}\delta_{\lambda,\overline{s}}\delta_{\lambda,-s},
\label{6}
\end{eqnarray}   
we find within the adopted accuracy
\begin{equation}
R_{s\overline{s}}^{(\lambda)}=\frac{1}{\sqrt{2}}
\left[\frac{m\omega}{\varepsilon\varepsilon'}
\delta_{\lambda,s}\delta_{\lambda,\overline{s}}-
\vartheta^{(\lambda)}\left(\delta_{\lambda,\overline{s}}\delta_{\lambda,-s} 
+\frac{\varepsilon}{\varepsilon'}
\delta_{\lambda,s}\delta_{\lambda,-\overline{s}}\right)\right]. 
\label{7}
\end{equation}
Taking advantage of relation $\delta_{\lambda, s}^2
=\delta_{\lambda, s}=(1+\xi)/2$, where $\xi=\lambda s$, we obtain
\begin{eqnarray}
&& R_{s\overline{s}}^{(\lambda)}(t_2)R_{s\overline{s}}^{(\lambda)\ast}(t_1)
=\frac{m^2}{8\varepsilon_{-}^2\varepsilon_{+}^2}
\Bigg\{\omega^2(1+\xi_{-})(1+\xi_{+})
\nonumber \\
&& +\gamma^2\vartheta^{(\lambda)}_2\vartheta^{(\lambda)\ast}_1
\left[\varepsilon_{+}^2(1-\xi_{-})(1+\xi_{+}) 
+\varepsilon_{-}^2(1+\xi_{-})(1-\xi_{+})\right]\Bigg\}. 
\label{8}
\end{eqnarray} 
Here for visualization we have written $\varepsilon=\varepsilon_-$
for the electron energy,
$\varepsilon'=\varepsilon_+$ for the positron energy,
$\xi_-=\lambda s, \xi_+=\lambda \overline{s}$ for description of
electron and positron polarization correspondingly,~
$\vartheta_{2,1} \equiv \vartheta(t_{2,1})$,
\begin{equation}
\vartheta^{(\lambda)}_2\vartheta^{(\lambda)\ast}_1=
(\vartheta_x(t_2)+i\lambda\vartheta_y(t_2))
(\vartheta_x(t_1)-i\lambda\vartheta_y(t_1))=
\mbox{\boldmath$\vartheta$}_2\mbox{\boldmath$\vartheta$}_1
-i\lambda(\textbf{n}(\mbox{\boldmath$\vartheta$}_2\times
\mbox{\boldmath$\vartheta$}_1)).
\label{9}
\end{equation}

Substituting Eqs.(\ref{8}), (\ref{9}) into Eq.(\ref{1}) one obtains
the completely differential probability of pair creation for circularly 
polarized photon and longitudinally polarized electron and positron. 
The above analysis shows that in the situation under consideration
the derivation with 
direct calculation of matrix element is very simple comparing
with the standard procedure (with calculation of traces for polarized
particles). This is in consequence of Eq.(\ref{6}).

It follows from Eqs.(\ref{7}), (\ref{8}) that in the case when the
energy of one of particles is essentially larger than other particle energy,
then together with energy the polarization of photon
is transmitted to this particle. This conclusion can be obtained 
immediately from Eqs.(\ref{3}), (\ref{4}) if one neglects the angle
$\mbox{\boldmath$\vartheta$}$ in the wave function Eq.(\ref{5}) 
for the particle with large energy and reserves it in the wave function
for the particle with small energy only.
Then either one wave function or another
becomes proportional to either $\varphi_s$ or $\varphi_{-\overline{s}}$.
In this case the polarization of particle (see Eq.(\ref{4}) is 
defined uniquely. If in this case in addition the angles 
$\vartheta \gg 1/\gamma$ contribute mainly ($\vartheta$ is the
angle between momenta of initial photon and created particle), then one 
can neglect the terms in Eqs.(\ref{7}), (\ref{8}) with the same polarizations
of created particles. Then in Eq.(\ref{8}) only one term remains
which defines uniquely the polarizations of both particles. This
fact agrees with helicity conservation rule in electromagnetic
processes. This situation is realized in the process of pair creation
by a photon in an external field for $\kappa \gg 1$ \cite{BKS},
as well as for the same process under strong influence of the LPM effect
\cite{BK3}.

Summing in Eq.(\ref{8}) over the polarizations of created electron $\xi_-$
and omitting the term 
$(\textbf{v}(\mbox{\boldmath$\vartheta$}_1\times
\mbox{\boldmath$\vartheta$}_2))$ which vanishes at integration
over the azimuthal angle of created pair we get
\begin{equation}
\sum_{\xi_{-}}R_{s\overline{s}}^{(\lambda)}(t_2)
R_{s\overline{s}}^{(\lambda)\ast}(t_1)
=\frac{m^2}{8\varepsilon_{-}^2\varepsilon_{+}^2}
\left\{\omega^2(1+\xi_{+})
+\gamma^2\mbox{\boldmath$\vartheta$}_2\mbox{\boldmath$\vartheta$}_1
\left[\varepsilon_{+}^2(1+\xi_{+}) 
+\varepsilon_{-}^2(1-\xi_{+})\right]\right\}. 
\label{9a}
\end{equation} 

The cross process of emission of photon with energy $\omega$ 
by an electron with  high energy $\varepsilon$ in 
oriented crystal was considered recently by authors \cite{BK4}.
It is evident that this process possesses the similar properties
in the case when the photon takes away an essential part of
electron energy ($\varepsilon'=\varepsilon - \omega \ll \varepsilon$).
The explicit expression for the transformed combination 
$R_{s\overline{s}}^{(\lambda)}(t_2)R_{s\overline{s}}^{(\lambda)\ast}(t_1)$
in Eq.(\ref{8}) can be found using the standard substitutions:
\begin{eqnarray}
&& \varepsilon_{-} \rightarrow \varepsilon',\quad
\varepsilon_{+} \rightarrow -\varepsilon,\quad
\omega \rightarrow -\omega,\quad
\lambda \rightarrow  -\lambda, 
\nonumber \\
&& \overline{s} \rightarrow -\zeta,\quad
s \rightarrow \zeta',\quad
\xi_{+} \rightarrow \xi,\quad
\xi_{-} \rightarrow -\xi'.
\label{10}
\end{eqnarray} 
As a result we find
\begin{eqnarray}
\hspace{-15mm}&& R_{\zeta'\zeta}^{(\lambda)\ast}(t_2)R_{\zeta'\zeta}^{(\lambda)}(t_1)
=\frac{m^2}{8\varepsilon'^2\varepsilon^2}
\Bigg\{\omega^2(1+\xi)(1-\xi')
\nonumber \\
\hspace{-15mm}&& +\gamma^2(\mbox{\boldmath$\vartheta$}_1\mbox{\boldmath$\vartheta$}_2+
i\lambda(\textbf{v}(\mbox{\boldmath$\vartheta$}_1\times
\mbox{\boldmath$\vartheta$}_2))
\left[\varepsilon^2(1+\xi)(1+\xi') 
+\varepsilon'^2(1-\xi)(1-\xi')\right]\Bigg\},
\label{11}
\end{eqnarray}
where $\xi=\lambda \zeta,~\xi'=\lambda \zeta'$.
Substituting Eq.(\ref{11}) into the formula for the probability of 
photon emission (see e.g.\cite{BKS}, p.63, Eq.(2.27))  one obtains
the completely differential probability of radiation of circularly 
polarized photon from longitudinally polarized electron in the case 
when the final electron has longitudinal polarization 
Summing in Eq.(\ref{11}) over polarizations of final electron $\zeta'$
and omitting the term 
$(\textbf{v}(\mbox{\boldmath$\vartheta$}_1\times
\mbox{\boldmath$\vartheta$}_2))$ which vanishes at integration
over angles of emitted photon we arrive to Eq.(1.8) of \cite{BK4}.

It should be noted that while the terms depending on angles in 
Eqs.(\ref{8}) and (\ref{11}) describe the spin correlations arising
from helicity conservation rule, the spin correlations
in terms $\propto \omega^2$ describes by rule of conservation of
projection of angular momentum on the direction of motion at 
zero angles. This explains absence of the term with 
$(1-\xi_{-})(1-\xi_{+})$ in Eq.(\ref{8}) and
the term with 
$(1-\xi)(1+\xi')$ in Eq.(\ref{11}).
 
It should be noted that a few different spin correlations are
known in an external field. But after averaging over directions of
crystal field only the longitudinal polarization considered here 
survives.

\section{General approach to pair creation in oriented crystal}
\setcounter{equation}{0}

The theory of high-energy electron radiation and electron-positron pair
creation in oriented crystals was developed in \cite{BKS1}-\cite{BKS2},
and given in \cite{BKS}. In these publications the process radiation
from electron and pair creation by a photon was considered  
for unpolarized particles. Since the expression for the pair creation
probability has the same structure as
Eqs.(\ref{8}), (\ref{9a}) below we use systematically the methods 
of mentioned papers to obtain the characteristics of pair creation
of longitudinally polarized electron (positron) by a circularly 
polarized photon.

Let $\vartheta_0$ be the photon angle of incidence with respect of
a chosen axis of crystal and $V_0$ be the scale of the corresponding 
continuous potential of the axis. For 
$\vartheta_0 \gg (V_0/\varepsilon)^{1/2} \equiv \vartheta_c$,
created particles are moving high above the potential barrier, and can
be described in terms of rectilinear trajectory. In rest frame of a 
crystal there is only an electric field of axis $\textbf{E}$. In frame
moving with a relativistic velocity $\textbf{v}=\textbf{n}v~(\textbf{n}=
\textbf{k}/\omega,~k^2=\omega^2-\textbf{k}^2=0)$ along the photon 
direction momentum, a magnetic field $\textbf{H}=\gamma_{v}(\textbf{E}
\times \textbf{v})$ arises $(\gamma_{v}=(1-v^2 )^{-1/2} \gg 1)$
and, as it is well known, that resultant field in this frame can be 
represented with relativistic accuracy in the form of plane waves with
frequencies $\gamma_{v}|q_{\parallel}|v(q_{\parallel}=\textbf{qn})$.

The periodic crystal potential $U(\textbf{r})$ can be
presented as the Fourier series (see e.g.\cite{BKS}, Sec.8 )
\begin{equation}
U(\textbf{r})=\sum_{\textbf{q}}G(\textbf{q})e^{-i\textbf{q}\textbf{r}},
\label{1.1}
\end{equation} 
where $\textbf{q}=2\pi(n_1, n_2, n_3)/l;~l$ is the lattice constant.

The equivalent photon flux averaged over time and over transverse 
coordinates is the sum of partial contributions $\textbf{J}_q$.
The later quantity has the form
\begin{equation}
\textbf{J}_q= -\textbf{n}\frac{\gamma_{v}}{4\pi e^2}
\frac{|G(\textbf{q})|^2}{|q_{\parallel}|}\textbf{q}_{\perp}^2 
\label{2.1}
\end{equation} 
where $\textbf{q}_{\perp}=\textbf{q}-\textbf{n(qn)}$. In the interaction 
region, the transverse size of which is of the order $\lambda_c=1/m$,
and the longitudinal size is the process formation length,
which is of the order $2\pi/\gamma_vq_{\parallel}$ in the center mass 
frame of the incident and equivalent photons, there are 
$N_q \simeq 2\pi \lambda_c^2 |\textbf{J}_q/\gamma_vq_{\parallel}|$
photons. The effective strength of interaction is characterized by the 
parameter
\begin{equation}
\alpha N_{ph} = \alpha \sum_{\textbf{q}} N_{\textbf{q}}
=\sum_{\textbf{q}} \frac{|G(\textbf{q})|^2}
{m^2|q_{\parallel}|^2}\textbf{q}_{\perp}^2 
\label{3.1}
\end{equation} 
This parameter is purely classical (it does not contain Planck's constant $\hbar$)
and always arises in problems with electromagnetic interaction in external field.
For $\alpha N_{ph} \ll 1$ the external field can be taken into account in
perturbation theory, while for $\alpha N_{ph} \gg 1$ one has the
constant field limit as it is known from the theory of interaction of photon
with the plane wave field (see e.g.\cite{BMS}).
For estimates one can assume $|G(\textbf{q})| \sim V_0,~q_{\parallel}
\sim q_{\perp}\vartheta_0$, in which case 
\begin{equation}
\alpha N_{ph} \sim \left(\frac{V_0}{m \vartheta_0}\right)^2.
\label{4.1}
\end{equation} 
Therefore for $\vartheta_0 \ll V_0/m \equiv \vartheta_v$ 
the constant field approximation
is applicable while for $\vartheta_0 \gg V_0/m$ the perturbation theory 
is valid, the first approximation of which is the coherent
pair production theory (see e.g. \cite{TM}).

In a crystal one have to integrate over pair creation points. Substituting
Eq.(\ref{9a}) into Eq.(\ref{1}) we find for the probability of creation
of longitudinally polarized positron by a circularly 
polarized photon
\begin{eqnarray}
&&dw_{\xi_+}=\frac{\alpha m^2}{(2\pi)^2\omega}
\frac{d^3p_+}{2\varepsilon_-\varepsilon_+}\int \frac{d^3r_0}{V}
\nonumber \\
&&\times \int dt_1\int dt_2 e^{iA_p}
\left[\varphi_{p1}(\xi_+)-\frac{1}{4}\gamma^2\left(\textbf{v}_1-
\textbf{v}_2\right)^2\varphi_{p2}(\xi_+)\right],
\nonumber \\
&&A_p=\frac{m^2 \omega}{2\varepsilon_+ \varepsilon_-}\int_{t_1}^{t_2}
\left[1+\gamma^2\mbox{\boldmath$\vartheta$}^2(t)\right]dt,
\nonumber \\
&&\varphi_{p1}(\xi_+)=1+\xi_+\frac{\omega}{\varepsilon_+}, \quad
\varphi_{p2}(\xi_+)=(1+\xi_+)\frac{\varepsilon_+}{\varepsilon_-}+
(1-\xi_+)\frac{\varepsilon_-}{\varepsilon_+},
\label{5.1}
\end{eqnarray} 
where $\alpha=e^2=1/137$, the vector $\mbox{\boldmath$\vartheta$}$ is defined 
in Eq.(\ref{5}), $V$ is the volume of the crystal, $\textbf{p}_+$
is the momentum of the positron in the creation point $\textbf{r}_0$.
Corresponding, the probability of creation
of longitudinally polarized electron by a circularly 
polarized photon is
\begin{eqnarray}
&&dw_{\xi_-}=\frac{\alpha m^2}{(2\pi)^2\omega}
\frac{d^3p_-}{2\varepsilon_-\varepsilon_+}\int \frac{d^3r_0}{V}
\nonumber \\
&&\times \int dt_1\int dt_2 e^{iA_p}
\left[\varphi_{e1}(\xi_-)-\frac{1}{4}\gamma^2\left(\textbf{v}_1-
\textbf{v}_2\right)^2\varphi_{e2}(\xi_-)\right],
\nonumber \\
&&\varphi_{e1}(\xi_-)=1+\xi_-\frac{\omega}{\varepsilon_-}, \quad
\varphi_{e2}(\xi_-)=(1-\xi_-)\frac{\varepsilon_+}{\varepsilon_-}+
(1+\xi_-)\frac{\varepsilon_-}{\varepsilon_+}.
\label{6.1}
\end{eqnarray} 

From Eq.(\ref{5.1}) one can find polarization of created positron
\begin{equation}
\mbox{\boldmath$\zeta$}_+=\zeta_p \lambda\textbf{v},\quad 
\zeta_p =\frac{dw_{\xi_+=1}-dw_{\xi_+=-1}}{dw_{\xi_+=1}+dw_{\xi_+=-1}}
\label{7.1}
\end{equation} 
In the limiting case $\varepsilon_+ \gg \varepsilon_-$ it follows from
Eq.(\ref{7.1}) that $\zeta_p \rightarrow 1$.

The particle velocity can be presented in a form
$\textbf{v}(t) =\textbf{v}_0+\Delta\textbf{v}(t)$, where 
$\textbf{v}_0$ is the average velocity. If 
$\vartheta_0 \gg \vartheta_c$, we find $\Delta\textbf{v}(t)$
using the rectilinear trajectory approximation for 
the potential Eq.(\ref{1.1})
\begin{equation}
\Delta\textbf{v}(t)=-\frac{1}{\varepsilon}\sum 
\frac{G(\textbf{q})}{q_{\parallel}}\textbf{q}_{\perp}
\exp [-i(q_{\parallel}t+\textbf{q}\textbf{r})],
\label{8.1}
\end{equation} 
where $ q_{\parallel}=(\textbf{q}\textbf{n}),~
\textbf{q}_{\perp}=\textbf{q}-\textbf{n}(\textbf{q}\textbf{n})$
(for detail see Eqs.(3.27)-(3.31) in \cite{BKS}).
Substituting Eq.(\ref{8.1}) into Eq.(\ref{5.1}) and performing 
integration over ${\bf u}={\bf n}-{\bf v}_0~(d^3p \simeq \varepsilon^2  
d\varepsilon d{\bf u})$ 
and passing to the variables $t, \tau:~t_1=t-\tau,~t_2=t+\tau$, 
we obtain after simple calculations the general expression for   
the probability of creation
of longitudinally polarized positron by a circularly 
polarized photon
\begin{eqnarray}
&& dW_{\xi_+} \equiv \frac{dw_{\xi_+}}{dt} 
= \frac{i\alpha m^2}{4 \pi\omega^2}d\varepsilon_+
\int \frac{d^3r_0}{V} 
\int \frac{d\tau}{\tau+i0}\Bigg[\varphi_{p1}(\xi_+)+\varphi_{p2}(\xi_+)
\nonumber \\
&& \times \left(\sum_{\textbf{q}} 
\frac{G(\textbf{q})}{m q_{\parallel}}\textbf{q}_{\perp}
\sin(q_{\parallel}\tau)e^{i\textbf{q}\textbf{r}_0} \right)^2 \Bigg]
  e^{iA_{p1}},
\label{9.1}
\end{eqnarray}
where 
\begin{eqnarray}
&& A_{p1}=\frac{m^2\omega\tau}{\varepsilon_-\varepsilon_+} \left[1+
\sum_{\textbf{q},\textbf{q}^{\prime} }\frac{G(\textbf{q})G(\textbf{q}')}
{m^2 q_{\parallel} q_{\parallel}'}
(\textbf{q}_{\perp}\textbf{q}_{\perp}')\Psi(q_{\parallel}, q_{\parallel}', \tau)
\exp [-i(\textbf{q}+\textbf{q}')\textbf{r}_0]\right] 
\nonumber \\
&& \Psi(q_{\parallel}, q_{\parallel}', \tau)=
\frac{\sin(q_{\parallel}+ q_{\parallel}')\tau}{(q_{\parallel}+ 
q_{\parallel}')\tau}-\frac{\sin(q_{\parallel}\tau)}{q_{\parallel}\tau}
\frac{\sin(q_{\parallel}'\tau)}{q_{\parallel}'\tau}.
\label{10.1}
\end{eqnarray}

\section{Pair creation for $\vartheta_0 \ll V_0/m$\\ 
(constant field limit and corrections to it)}
\setcounter{equation}{0}

The behavior of probability  Eq.(\ref{9.1}) for various entry angles and
energies is determined by the dependence on these parameters of the phase $A_{p1}$ 
given Eq.(\ref{10.1}). Here we consider the axial case for
$\vartheta_0 \ll V_0/m \equiv \vartheta_v$. The direction of crystal axis 
we take as $z-$axis of the coordinate system. The order of magnitude of 
the double sum in 
$A_{p1}$ is $(G/m)^2(q_{\perp}/q_{\parallel})^2\Psi(q_{\parallel}, q_{\parallel}', \tau)$.
For the vector $\textbf{q}$ lying in the plane $(x,y)$ we introduce notation
$\textbf{q}_t$, for such vectors one has $q_z=0$ and the quantities in 
Eq.(\ref{10.1}) can be estimated in the following way:
\begin{equation}
G(\textbf{q}) \sim V_0,\quad q_{\perp} \sim 1/a,\quad q_{\parallel} 
\sim \vartheta_0/a,
\label{1.2}
\end{equation}
where $a$ is the size of the region of action of the continuous potential. For all 
remaining vectors $q_{\perp} \sim q_{\parallel} \sim 1/a$. Then the contribution 
to the sum of the terms with $q_z \neq 0$ will be $\sim (V_0/m)^2\Psi \leq (V_0/m)^2$.
Since $(V_0/m)^2 \ll 1$ this contribution can be neglected. Thus, we keep in the sum
only terms with $\textbf{q}_t$ for which its value is $\sim (V_0/m\vartheta_0)^2\Psi$.
The large value of the phase $A_{p1}$ leads to an 
exponential suppression of probability
$dW_{\xi_+}$. Therefore the characteristic value of the variable $\tau$ in the integral
Eq.(\ref{10.1}) (which have the meaning of the formation time (length) of the process)
will be adjusted in a such way that the large factor $(V_0/m\vartheta_0)^2$ will be
compensated by the function $\Psi(q_{\parallel}, q_{\parallel}', \tau)$, i.e. for
small entry angles the contribution gives region where $q_{\parallel}\tau \ll 1$.
Expanding the phase $A_{p1}$ correspondingly we find an approximate expression for
$\vartheta_0 \ll \vartheta_v$ 
\begin{eqnarray}
&& A_{p1} \simeq \frac{m^2\omega\tau}{\varepsilon_-\varepsilon_+}\Bigg\{1-
\frac{\tau^2}{3} \sum_{\textbf{q}_t,\textbf{q}_t'}
G(\textbf{q}_t)G(\textbf{q}_t')\frac{(\textbf{q}_t\textbf{q}_t')}{m^2}
\exp\left[-i(\textbf{q}_t+\textbf{q}_t')\mbox{\boldmath$\varrho$} \right]
\nonumber \\
&& \times \left[ 1-\frac{\tau^2}{10}\left((\textbf{n}\textbf{q}_t)^2
+ (\textbf{n}\textbf{q}_t')^2
+\frac{2}{3} (\textbf{n}\textbf{q}_t)
(\textbf{n}\textbf{q}_t')\right) \right]\Bigg\},
\label{2.2}  
\end{eqnarray} 
here $\mbox{\boldmath$\varrho$}=\textbf{r}_{0t}$. We can rewrite 
Eq.(\ref{2.2}) in the terms of the average potential of atomic string
$\displaystyle{U(\mbox{\boldmath$\varrho$})=\sum_{\textbf{q}_t} G(\textbf{q}_t)
\exp(-i\textbf{q}_t \mbox{\boldmath$\varrho$})}$:
\begin{equation}
A_{p1} = \frac{m^2\omega\tau}{\varepsilon_-\varepsilon_+}
\left\lbrace 1+\frac{\tau^2}{3} \textbf{b}^2\tau^2 +
 \frac{\tau^4}{15} \left[(\textbf{b}(\textbf{n}
\mbox{\boldmath$\nabla$})^2 \textbf{b}) 
+\frac{1}{3} ((\textbf{n}
\mbox{\boldmath$\nabla$})\textbf{b})^2 \right]  \right\rbrace, 
\label{3.2}
\end{equation} 
where $\textbf{b}=\mbox{\boldmath$\nabla$}U(\mbox{\boldmath$\varrho$})/m,~
\mbox{\boldmath$\nabla$}=\partial/\partial \mbox{\boldmath$\varrho$}$.
For the pre-exponential factor in Eq.(\ref{9.1}) we find 
\begin{equation}
[\ldots] \simeq \varphi_{p1}(\xi_+) - \varphi_{p2}(\xi_+)\tau^2\left[\textbf{b}^2 +
\frac{\tau^2}{3} (\textbf{b}(\textbf{n}
\mbox{\boldmath$\nabla$}) \textbf{b}) \right] 
\label{4.2}
\end{equation}

Taking the integral over $\tau$ we obtain the spectral probability
for $\vartheta_0 \ll V_0/m$.
\begin{eqnarray}
&& dW^F_{\xi_+}(\varepsilon_+)=\frac{\alpha m^2d\varepsilon_+}{2\sqrt{3}\pi\omega^2}
\int \frac{d^2\varrho}{S} \Bigg\{R_0(\lambda)
 -\frac{(\textbf{b}(\textbf{n}
\mbox{\boldmath$\nabla$})^2 \textbf{b})}{3\textbf{b}^4}
R_1(\lambda)
\nonumber \\
&&-\frac{\lambda}{30\textbf{b}^4}\left[((\textbf{n}
\mbox{\boldmath$\nabla$})\textbf{b})^2 +3 (\textbf{b}(\textbf{n}
\mbox{\boldmath$\nabla$})^2 \textbf{b})\right]R_2(\lambda)\Bigg\}, 
\label{5.2}
\end{eqnarray} 
where 
\begin{eqnarray}
\hspace{-15mm}&& R_0(\lambda)=\varphi_{p2}(\xi_+)K_{2/3}(\lambda)+
\varphi_{p1}(\xi_+)\int_{\lambda}^{\infty} K_{1/3}(y)dy, 
\nonumber \\
&&R_1(\lambda)=
\varphi_{p2}(\xi_+)\left(K_{2/3}(\lambda)-\frac{2}{3\lambda}K_{1/3}(\lambda)
\right),
\nonumber \\
\hspace{-15mm}&& R_2(\lambda)=\varphi_{p1}(\xi_+)\left( K_{1/3}(\lambda) 
-\frac{4}{3\lambda}K_{2/3}(\lambda)\right)
\nonumber \\
&&-\varphi_{p2}(\xi_+)\left(\frac{4}{\lambda}K_{2/3}(\lambda) 
-\left(1+\frac{16}{9\lambda^2} \right)K_{1/3}(\lambda) \right),
\label{6.2}
\end{eqnarray} 
here $\displaystyle{\lambda=\frac{2m^2\omega}{3\varepsilon_-\varepsilon_+|\textbf{b}|}}$,
$K_{\nu}(\lambda)$ is the modified Bessel function (McDonald's function). 
Since the expression for $dW^F_{\xi_+}$ is independent of $z$, it follows that 
$\int d^3r_0/V \rightarrow \int d^2\varrho/s$, where $S$ 
is the transverse cross
section area per axis. The term in Eq.(\ref{5.2}) with $R_0(\lambda)$
represent the spectral probability in the constant field limit.
The other terms are the correction proportional $\vartheta_0^2$
arising due to nongomogeneity of field in crystal.

If the potential $U(\mbox{\boldmath$\varrho$})$ can be assumed to be
axially symmetric, we put  $U=U(\mbox{\boldmath$\varrho$}^2)$ and one 
can integrate over angles of vector $\mbox{\boldmath$\varrho$}$. We
obtain
\begin{eqnarray}
&& dW^F_{\xi_+}(\varepsilon_+) =\frac{\alpha m^2 d\varepsilon_+}{2\sqrt{3}\pi\omega^2}
\int_{0}^{x_0} \frac{dx}{x_0} \Bigg\{ R_0(\lambda)
 -\frac{1}{6}\left( \frac{m \vartheta_0}{V_0}\right)^2\Bigg[
\frac{xg''+2g'}{xg^3} R_1(\lambda)
\nonumber \\
&& -\frac{\lambda}{20g^4x^2}\left(2x^2g^{\prime 2}+g^2+14gg^{\prime}x
+6x^2gg^{\prime \prime}\right)R_2(\lambda)\Bigg] \Bigg\}, 
\label{7.2}
\end{eqnarray} 
where we have gone over to the new variable $x=\mbox{\boldmath$\varrho$}^2/a_s^2,
~x \leq x_0,~x_0^{-1}=\pi a_s^2dn_a=\pi a_s^2/S,~a_s$ 
is the effective screening radius
of the potential of the string, $n_a$ is the density of atoms in a crystal,
$d$ is the average distance between atoms of a chain forming the axis.
The notation $U^{\prime}(x)=-V_0g(x)$ is used in Eq.(\ref{7.2}) and
\begin{equation}
\lambda=
\frac{2\omega^2}{3\varepsilon_-\varepsilon_+\kappa(x)},~
\kappa_s=\frac{V_0\omega}{m^3 a_s},
~\kappa(x)=-\frac{dU}{d\varrho}\frac{\omega}{m^3}=2\kappa_s\sqrt{x}g(x).
\label{9.2}
\end{equation} 
For specific calculation we use the following expression for the potential of axis:
\begin{equation}
U(x)=V_0\left[\ln\left(1+\frac{1}{x+\eta} \right)- 
\ln\left(1+\frac{1}{x_0+\eta} \right) \right],~g(x)=\frac{1}{(x+\eta)(x+\eta+1)}. 
\label{10.2}
\end{equation} 
For estimates one can put $V_0 \simeq Ze^2/d,~\eta \simeq 2u_1^2/a_s^2 \equiv \eta_1$,
where $Z$ is the charge of the nucleus, 
$u_1$ is the amplitude of thermal vibrations,
but actually the parameters of potential were determined by means of a fitting
procedure using the potential Eq.(\ref{1.1}) (table of parameters for 
different crystals is given in Sec.9 of \cite{BKS}).
The function $\kappa(x)$ vanishes at $x=0$ as a result of
thermal vibrations and reaches a maximum at 
\begin{equation}
x=x_m=\frac{1}{6}\left\lbrace \left[1+16\eta(1+\eta)\right]^{1/2}-1-2\eta \right\rbrace 
\label{12.2}
\end{equation} 
and then fall off as $x^{-3/2}$. It should be noted, that for all 
crystals $\eta \ll 1$ and in this case $x_m \simeq \eta$ and 
$\kappa(x_m)\equiv \kappa_m \simeq \kappa_s/\sqrt{\eta}$.

In Fig.1 the spectral probability of coherent pair creation
$dw_{\xi_+}/dy$ in tungsten, T=293 K, axis $<111>$ is given. 
The calculation was performed using Eq.(\ref{7.2}) with term 
$R_0(\lambda)$ only (the corrections $\propto \vartheta_0^2$ 
were omitted). The sum of curves at the indicated energy 
gives unpolarized case. At energies $\omega=100~$GeV and
$\omega=250~$GeV the coherent process dominates over the
Bethe-Heitler mechanism.

Let us consider the case of small values of parameter $\kappa$. In the constant 
field approximation (the term with $R_0(\lambda)$ in 
\textsc{Eq.(\ref{7.2}))} in the probability $dW^F_{\xi_+}(\varepsilon_+)$ 
one can substitute the asymptotic of McDonald's functions
$K_{\nu}(\lambda)$ at $\lambda \gg 1$. After this one obtains using 
the Laplace method in 
integration over the transverse coordinate $x$
\begin{equation}
 \frac{dW^F_{\xi_+}(\varepsilon_+)}{d\varepsilon_+}=
\frac{\sqrt{3}\alpha m^2}{4 \omega^2x_0}\left[\frac{\varepsilon_+}{\omega}(1+\xi_+) 
+\frac{\varepsilon_-^2}{\omega^2} \right] 
\frac{\kappa_m^{3/2}}{\sqrt{-\kappa_m''}} 
\exp\left(-\frac{2\omega^2}{3\varepsilon_-\varepsilon_+\kappa_m} \right),
\label{13.2}
\end{equation}  
where $\kappa_m=\kappa(x_m),~\kappa_m''=\kappa''(x_m)$. 
For unpolarized particle this result
coincides with Eq.(12.14) of \cite{BKS}.
If $\kappa_m \ll 1$ one can integrate Eq.(\ref{13.2}) also using the Laplace method.
The integral probability of polarized pair creation is
\begin{equation}
W^F_{\xi_+} = \frac{9\alpha m^2}{64x_0\omega}\sqrt{\frac{\pi}{2}}
\frac{\kappa_m^{2}}{\sqrt{-\kappa_m''}}
\exp\left(-\frac{8}{3\kappa_m} \right)
\left(1+\frac{2}{3} \xi_+\right).
\label{14.2} 
\end{equation} 

At relatively low photon energies where the parameter $\kappa_m$
is small
\begin{equation}
\kappa_m \simeq \frac{\kappa_s}{\sqrt{\eta}}\simeq \frac{\kappa_s}{\sqrt{\eta_1}}
=\frac{\omega V_0}{\sqrt{2}m^3 u_1} \equiv \frac{\kappa_1}{\sqrt{2}}
\label{1.3} 
\end{equation} 
and the main contribution to  $e^+e^-$ pair creation gives the incoherent 
(Bethe-Heitler) mechanism. The influence of effective crystalline
fields on incoherent processes is discussed in \cite{BKS}, Sections
7, 21, but the polarization effects were not included.

In the case where the influence of effective crystalline
field is weak ($\kappa_1 \ll 1$), the cross section acquires
the correction of the order  $\kappa_1^2$ (we proceed as in Sec.21.4 in 
\cite{BKS} but taking into account polarization):
\begin{equation}
\sigma_p(\xi_+)=\frac{1}{2}\left(\sigma_p+ \xi_+\sigma_{\xi_+}\right), 
\label{2.3}
\end{equation} 
where
\begin{eqnarray}
&& \sigma_p=\frac{28 Z^2\alpha^3}{9m^2}
\left[\left(1+\frac{396}{1225}
\overline{\kappa^2}\right)L_u-\frac{1}{42}- 
\frac{1789}{6125} \overline{\kappa^2} \right],
\nonumber\\
&& \sigma_{\xi_+}=\frac{4Z^2\alpha^3}{3m^2}
\left[\left(1+\frac{32}{75}
\overline{\kappa^2}\right)L_u-\frac{1}{6}- 
\frac{439}{1125} \overline{\kappa^2} \right],
\nonumber\\
&& L_u=L_0-h\left(\frac{u_1^2}{a^2}\right),\quad L_0=\ln(ma)+\frac{1}{2}-f(Z\alpha),
\quad a=111Z^{-1/3}\lambda_c. 
\label{3.3} 
\end{eqnarray}
Here $\lambda_c$ is the electron Compton wavelength, 
$\overline{\kappa^2}$ is the mean value of 
$\kappa^2(\mbox{\boldmath$\varrho$})$ for the atomic density
$n_a(\mbox{\boldmath$\varrho$})=
\exp(-\mbox{\boldmath$\varrho$}^2/2u_1^2)/2\pi u_1^2$,
the function $h((u_1^2/a^2)$ reflects a nongomogeneity
of atomic distribution in crystal, the function $f(Z\alpha)$
represents the Coulomb corrections:
\begin{eqnarray}
&&\overline{\kappa^2}=\int \kappa^2(\mbox{\boldmath$\varrho$})
n_a(\mbox{\boldmath$\varrho$})d^2\varrho=\int_{0}^{\infty}
\frac{e^{-x/\eta_1}}{\eta_1}\kappa^2(x)dx,~ \eta_1=\frac{2u_1^2}{a_s^2}
\nonumber\\
&& f(\xi)={\rm Re}\left[\psi(1+i\xi)-\psi(1) \right]
=\xi^2\sum_{n=1}^{\infty} \frac{1}{n(n^2+\xi^2)}, 
\label{4.3} 
\end{eqnarray} 
where $\psi(\xi)$ is the logarithmic derivative of the
gamma function, Ei($-x$) is the exponential integral function.
It is seen from Eq.(\ref{3.3}) that in a weak field limit 
($\kappa_1 \ll 1$) in the logarithmic approximation
the integral positron polarization is $\zeta_{+}=\zeta^C \simeq 3/7$. 

The probability of coherent pair creation becomes comparable
with the Bethe-Maximon probability at $\kappa_m \sim 1$.
In this situation Eq.(\ref{14.2}) still has quite satisfactory accuracy
(error $< 15\%$), while Eq.(\ref{3.3}) becomes inapplicable.
The relative contribution of the LPM effect to the total probability 
of pair creation by a photon was found recently in \cite{BK5}.
This contribution has the maximum about 5\% at $\kappa_m \sim 1$
for heavy elements. In this paper we neglect the LPM effect.
The differential cross section of incoherent pair creation in
the effective crystalline field for polarized particles in the 
logarithmic approximation for an arbitrary photon energy can be 
presented in a form
\begin{equation}
\frac{d\sigma_p(\xi_+)}{dy}=\frac{2 Z^2\alpha^3}{15m^2}L(\kappa_1)
\int_{0}^{\infty}\exp\left(-\frac{x}{\eta_1}\right)f(x, y, \xi_+)\frac{dx}{\eta_1},
\quad L(\kappa_1)=L_u+\frac{\ln(1+\kappa_1)}{3},
\label{5.3}
\end{equation} 
where 
\begin{equation}
f(x, y, \xi_+)=(1+\xi_+)f_1(x, y) + \left[y^2 (1+\xi_+)
+(1-y)^2(1-\xi_+)\right]f_2(x, y). 
\label{6.3}
\end{equation} 
Here
\begin{eqnarray}
\hspace{-10mm}&& f_1(x, y)=z^4\Upsilon(z)-3z^2\Upsilon'(z)-z^3,\quad
f_2(x, y)=(z^4+3z)\Upsilon(z)-5z^2\Upsilon'(z)-z^3,
\nonumber \\
\hspace{-10mm}&& z=z(x, y)=\left[\kappa(x)y(1-y) \right]^{-2/3},\quad
y=\frac{\varepsilon_+}{\omega},
\label{7.3}
\end{eqnarray} 
where the function $\kappa(x)$ is defined in Eq.(\ref{9.2}).
Here $\Upsilon(z)$ is the Hardy function
\begin{equation}
\Upsilon(z)=\int_{0}^{\infty}\sin\left(z\tau+\frac{\tau^3}{3} \right). 
\label{8.3}
\end{equation} 
When the influence of axis field on the incoherent process is weak
($\kappa_1 \ll 1, z \gg 1$), one can use decomposition (see Appendix D, 
Eq.(\ref{c.3}))
\begin{equation}
\Upsilon(z) \simeq \frac{1}{z}+\frac{2}{z^4}+\frac{40}{z^7},\quad
f_1(z) \simeq 5+\frac{64}{z^3}, \quad f_2(z) \simeq 10+\frac{86}{z^3}.
\label{9.3}
\end{equation} 
Substituting this decomposition in Eq.(\ref{5.3}) and integrating
over $y$ we obtain within the logarithmic accuracy Eq.(\ref{3.3}).

Integrating Eq.(\ref{5.3}) over the positron energy we find for the
cross sections contained in Eq.(\ref{2.3})
\begin{eqnarray}
\hspace{-10mm}&& \sigma_p=\frac{4 Z^2\alpha^3}{15m^2} L_p(\kappa_m)
\int_{0}^{1}dy\int_{0}^{\infty}\exp\left(-\frac{x}{\eta_1}\right)
\left[f_1(x,y) + 2y^2 f_2(x,y)\right] \frac{dx}{\eta_1}, 
\nonumber \\
\hspace{-10mm}&& \sigma_{\xi_+}=\frac{4 Z^2\alpha^3}{15m^2} 
 L_{\xi_+}(\kappa_m)
\int_{0}^{1}dy\int_{0}^{\infty}\exp\left(-\frac{x}{\eta_1}\right)
f_1(x,y) \frac{dx}{\eta_1}
\label{9.3a}
\end{eqnarray} 
Here we introduced the functions $L_p(\kappa_m)$ and $L_{\xi_+}(\kappa_m)$
which refined the logarithmic approximation 
Eq.(\ref{5.3}) for $\kappa_1 \leq 1$ and for $\kappa_1 \gg 1$.
These functions are obtained by means of interpolation procedure
(\cite{BK5}):
\begin{eqnarray}
\hspace{-10mm}&& L_p(\kappa_m)=L_u-\frac{1}{42}+\frac{1}{3}
\ln \frac{6-3\kappa_m^2+3\kappa_m^3}{6+\kappa_m^2},
\nonumber \\
\hspace{-10mm}&& L_{\xi_+}(\kappa_m)=L_u-\frac{1}{6}+\frac{1}{3}
\ln \frac{9-6\kappa_m^2+6\kappa_m^3}{9+2\kappa_m^2}
\label{9.3b}
\end{eqnarray} 
In Fig.2 the integral probability of pair creation by a photon in
tungsten, axis $<111>$, $T=100~$ K as function of photon energy
is shown. The curve 1 is
the incoherent and the curve 2 is the coherent contribution while the curve 3
is their sum giving the total probability. In low energy region
the LPM effect diminishes slightly the total probability (see \cite{BK5}).
In this region the incoherent contribution dominates, these contributions
are equal at $\omega \simeq 10~$GeV. At higher energies 
the coherent contribution dominates achieving maximum 
at $\omega \simeq 1.2~$TeV, while
the incoherent contribution aims for zero.

In Fig.3 the positron polarization vs relative positron energy 
$\varepsilon/\omega$ is given: the curve 1 for $\omega=12~$GeV,
the curve 2 for $\omega=22~$GeV, the curve 3 for $\omega=100~$GeV.
Both coherent and incoherent contributions are taken into account.
It is seen that shown dependence is approximately universal. When
$\varepsilon/\omega \rightarrow 0$ the positron polarization is tending
to the Coulomb limit $\zeta^C=-1/3$ since in this limit the coherent  
contribution is negligible. In the limit $\varepsilon/\omega 
\rightarrow 1$ the polarization $\xi_+ \rightarrow 1$
because of helicity transfer, which is common
for both coherent and incoherent contributions.

In Fig.4 the integral positron polarization $\xi_+$ as a function of 
photon energy $\omega$ is shown. 
Both coherent and incoherent contributions are taken into account.
At $\omega \rightarrow 0$ this is
is the Coulomb limit $\zeta^C=3(1-1/7L_u)/7$ ($L_u$ is defined in
Eq.(\ref{3.3})). At intermediate energies the shown dependence is the
result of interplay of the coherent and the incoherent contributions, while
in the high energy region we see the behavior of the coherent contribution.

Now we turn to the situation when the photon energy is very high and
influence of crystalline field becomes very strong
($\kappa_1 \gg 1$).
In this case we have (see Appendix B) for Eq.(\ref{2.3})
($\sigma_p$ is given in Eq.(21.34) in \cite{BKS})
\begin{eqnarray}
&&\sigma_p \simeq \frac{8 Z^2\alpha^3}{25m^2}
\frac{\Gamma^3(1/3)}{}3^{2/3}\Gamma(2/3) \overline{\kappa^{-2/3}}L_p(\kappa_m)
\simeq 3.23 
\frac{Z^2\alpha^3 L(\kappa_1)}{m^2 \kappa_1^{2/3}},
\nonumber \\
&&\sigma_{\xi_+} \simeq \frac{1.54 Z^2\alpha^3}{m^2}\overline{\kappa^{-4/3}}L_{\xi_+}(\kappa_m)
\simeq 3.95 
\frac{Z^2\alpha^3 L(\kappa_1)}{m^2 \kappa_1^{4/3}},
\label{10.3}
\end{eqnarray} 
where the average $\overline{\kappa^{s}}$ is defined in Eq.(\ref{4.3}).
It should be noted that the asymptotic expansion of $\sigma_{\xi_+}$
at $\kappa_1 \gg 1$ becomes valid only for very large values of
$\kappa_1$ when the ratio $\sigma_{\xi_+}/\sigma_p$ is very small.
This is the consequence of specific form of the next term of
decomposition of $\sigma_{\xi_+}= \sigma_{\xi_+}^{(1)}+
\sigma_{\xi_+}^{(2)}+\ldots$~.
The term $\sigma_{\xi_+}^{(1)}$ (Eq.(\ref{10.3})) is calculated 
in Appendix C. The next term
$\sigma_{\xi_+}^{(2)}$ has a form
\begin{equation}
\sigma_{\xi_+}^{(2)}=\frac{Z^2\alpha^3}{m^2}
\frac{A(\ln \kappa_1+B)}{\kappa_1^2}L_{\xi_+}(\kappa_m).
\label{10.3a}
\end{equation}
For example, for tungsten crystal, axis $<111>$, T= 100 K,
the obtained in numerical calculation values of the coefficients 
are $A \simeq 10,~B \simeq -1.7$.

In the region $\kappa_s \gg 1$ (the constant field
approximation is valid 
if $\kappa_s^{1/3}\vartheta_0/\vartheta_v \ll 1$) 
the integral probability of coherent pair creation 
(the terms $\propto \xi_+$ are calculated in 
Appendix B , the integral probability for the unpolarized case is given 
in Eq.(12.16) of \cite{BKS}) is
\begin{eqnarray}
&& W_p^F(\xi_+)= c_0\frac{\alpha V_0}{mx_0a_s\kappa_s^{1/3}}
\Big[\ln \kappa_s +B_1(\eta)
\nonumber \\
&& +\frac{21}{20}\xi_+\left(1-\frac{c_1}{\kappa_s^{2/3}}
(\ln \kappa_s+c_2) \right)\Big],  
\label{11.3}
\end{eqnarray} 
where (see Appendix B, Eq.(\ref{b.5}))
\begin{eqnarray}
\hspace{-10mm}&&c_0=\frac{5}{7\pi}
\frac{3^{1/6}\Gamma^3(2/3)}{2^{1/3}\Gamma(1/3)}
\simeq 0.201, \quad c_1=\frac{2^{7/3}\pi}{3^{19/6}}
\frac{\Gamma(1/3)}{\Gamma^3(2/3)} \simeq 0.527,
\nonumber \\
\hspace{-10mm}&&c_2=\ln 2-\frac{\ln 3}{2} -C+\frac{3}{2} \simeq 1.067;
\quad \beta=\frac{\eta}{1+\eta}
\nonumber \\
\hspace{-10mm}&& B_1(\eta)=-0.374-3.975\beta^{2/3}\left(1+\frac{8}{15}\beta+
\frac{7}{18}\beta^2 \right) + \beta\left(\frac{3}{2}+\frac{9}{8}\beta+
\frac{13}{14}\beta^2 \right). 
\label{12.3}
\end{eqnarray} 
When $\kappa_s=u_1 \kappa_1/a_s \gg 1$ the contribution of incoherent
process is very small and can be neglected
(see Eq.(21.35) of \cite{BKS}):
\begin{equation}
\frac{W_p^{incoh}}{W_p^F} \leq 10^{-2}
\frac{Z\alpha \ln(mu_1 \kappa_1^{1/3})}{\kappa_1^{1/3} \ln \kappa_s}.
\label{13.3}
\end{equation} 

\section{Modified theory of coherent pair production}
\setcounter{equation}{0}

The estimates of double sum in the phase $A_{p1}$ made at the beginning 
of previous section: $\sim (\vartheta_v/\vartheta_0)^2\Psi$ remain valid also for 
$\vartheta_0 \geq \vartheta_v$, except that now the factor in the double sum is
$(\vartheta_v/\vartheta_0)^2 \leq 1$, so that the values 
$|q_{\parallel}\tau| \geq 1$ contribute. We consider first the limiting case
$\vartheta_0 \gg \vartheta_v$, then this factor is small and $\exp(iA_{p1})$
can be expanded accordingly. As a result Eq.(\ref{9.1}) acquires the
form
\begin{eqnarray}
\hspace{-12mm}&& dW_{\xi_+}^{coh}(\omega)=\frac{i\alpha m^2d\varepsilon_+}{4\pi\omega^2}
\int_{-\infty}^{\infty} \frac{d\tau}{\tau+i0} 
\exp\left(i\frac{m^2\omega\tau}{\varepsilon\varepsilon'} \right)
\sum_{\textbf{q},\textbf{q}^{\prime} }\frac{G(\textbf{q})G(\textbf{q}')}
{m^2 q_{\parallel} q_{\parallel}'}
(\textbf{q}_{\perp}\textbf{q}_{\perp}')
\nonumber \\
\hspace{-12mm}&& \times\left[\varphi_{p2}(\xi_+)\sin(q_{\parallel}\tau) \sin(q_{\parallel}'\tau)
+i\varphi_{p1}(\xi_+)\frac{m^2\omega\tau}{\varepsilon_-\varepsilon_+} 
\Psi(q_{\parallel}, q_{\parallel}', \tau)\right] 
\nonumber \\
\hspace{-12mm}&&\times\int \frac{d^3r}{V} \exp [-i(\textbf{q}+\textbf{q}')\textbf{r}].
\label{1.4}
\end{eqnarray} 
The integration over coordinate $\textbf{r}$ in Eq.(\ref{1.4}) 
is elementary and
gives $\delta_{\textbf{q}+\textbf{q}', 0}$, after which the sum over $\textbf{q}'$
and the integrals over $\tau$ are easily calculated by means of the theory of
residues. Finally we obtain
\begin{eqnarray}
&& dW_{\xi_+}^{coh}(\omega)=\frac{\alpha d\varepsilon_+}{8\omega^2}
\sum_{\textbf{q}} |G(\textbf{q})|^2 \frac{\textbf{q}_{\perp}^2}{q_{\parallel}^2}
\left[\varphi_{p2}(\xi_+) +\varphi_{p1}(\xi_+)\frac{2m^2\omega}
{\varepsilon_-\varepsilon_+q_{\parallel}^2}\left(|q_{\parallel}|- 
\frac{m^2\omega}{2\varepsilon_-\varepsilon_+}\right)  \right] 
\nonumber \\
&& \times \vartheta\left(|q_{\parallel}|- 
\frac{m^2\omega}{2\varepsilon_-\varepsilon_+}\right). 
\label{2.4}
\end{eqnarray} 
For unpolarized photons ($\xi_+=0$) Eq.(\ref{2.4}) coincides with the result 
of standard theory of coherent pair production (CPP), see e.g. \cite{TM}. 

In the case $\kappa_s \gg 1$ ($\kappa_s$ is defined in Eq.(\ref{9.2})), one can
obtain from general expression Eq.(\ref{10.1}) the expression for spectral
distribution, the region of applicability of which is broader than that of
standard CPP theory. For this purpose it is necessary to take into account that
the phase $A_{p1}$ Eq.(\ref{10.1}) has for $q_{\parallel}+ q_{\parallel}' \neq 0$
terms of the order $(\vartheta_v/\vartheta_0)^3/\kappa'_s$ and 
$(\vartheta_v/\vartheta_0)^4/\kappa_s^{'2}~
(\kappa_s'=\kappa_s \varepsilon_-\varepsilon_+/\omega^2)$ 
which can be small even for 
$\vartheta_0 \leq\vartheta_v$ if $\kappa_s \gg 1$. Therefore, assuming that 
these contributions are small, we carry out the corresponding expansion of $\exp(iA_{p1})$,
while the term with $q_{\parallel}+ q_{\parallel}' = 0$ in the double sum in 
$A_{p1}$ will be retained in the exponent. As a result we obtain an expression 
which coincides in the form with Eq.(\ref{1.4}) where we must make the 
substitution
\begin{equation}
\exp\left(i\frac{m^2\omega\tau}{\varepsilon_-\varepsilon_+} \right) 
\rightarrow
\exp\left(i\frac{m_{\ast}^2\omega\tau}{\varepsilon_-\varepsilon_+}\right), 
\quad
m_{\ast}^2=m^2\left(1+\frac{\varrho}{2} \right). 
\label{3.4}
\end{equation} 
Above the parameter $\varrho$ (Eq.(\ref{1.1})) has the form
\begin{equation}
\frac{\varrho}{2}=\frac{1}{m^2}\sum_{\textbf{q},\textbf{q}'}G(\textbf{q})
G(\textbf{q}')
\frac{\textbf{q}_{\perp}\textbf{q}_{\perp}'}{q_{\parallel} q_{\parallel}'}
\left[\delta_{q_{\parallel}+ q_{\parallel}',0} -\delta_{q_{\parallel},0}
\delta_{q_{\parallel}',0}\right] =\sum_{\textbf{q},q_{\parallel} \neq 0}
\frac{|G(\textbf{q})|^2\textbf{q}_{\perp}^2 }{m^2q_{\parallel}^2},
\label{4.4}
\end{equation} 
and in the term $\displaystyle{\frac{\sin(q_{\parallel}+ q_{\parallel}')\tau}
{(q_{\parallel}+ q_{\parallel}')\tau}}$ it is necessary to assume that
$q_{\parallel}+ q_{\parallel}' \neq 0$. The remaining calculations 
are carried out in the same way as in the transition from Eq.(\ref{1.4}) to
Eq.(\ref{2.4}). The final result can be presented in a form 
\begin{eqnarray}
&& dW_{\xi_+}^{mcoh}=\frac{\alpha dy}{8\omega}
\sum_{\textbf{q}} |G(\textbf{q})|^2 \frac{\textbf{q}_{\perp}^2}{q_{\parallel}^2}
\Bigg[\frac{y}{1-y}(1+\xi_+) + \frac{1-y}{y} (1-\xi_+)
\nonumber \\
&& +\frac{8(y+\xi_+)}{(2+\varrho)y}\frac{\tau}{\tau_0}\left(1- 
\frac{\tau}{\tau_0}\right) \Bigg] \vartheta(\tau_0-\tau), 
\label{5.4}
\end{eqnarray}
where
\begin{equation}
\tau_0=\frac{4\omega |q_{\parallel}|}{m^2 (2+\varrho)}, \quad
\tau=\frac{1}{y(1-y)},\quad \tau_{min}=4.
\label{5.4a}
\end{equation}  
In derivation of Eq.(\ref{5.4}) the higher order terms over 
$\tau\varrho/\tau_0(2+\varrho)$ are omitted. 
Equation (\ref{5.4}) is not more complicated than Eq.(\ref{2.4})
but has a significantly broader range of applicability.

The spectral distributions Eqs.(\ref{2.4}) and (\ref{5.4}) can be
much higher than the Bethe-Heitler pair production distribution for 
small angles of incidence $\vartheta_0$ with respect to selected axis.
For the case $\vartheta_0 \ll 1$ the quantity $q_{\parallel}$
can be represented as
\begin{equation}
q_{\parallel} \simeq \frac{2\pi}{d}m+\textbf{q}_{\perp}\textbf{n}_{\perp}.  
\label{5.4b}
\end{equation} 
The main contribution to the sum in Eqs.(\ref{2.4}) and (\ref{5.4}) for
small $\vartheta_0$ is given $\textbf{q}$ with $m=0$, then 
\begin{equation}
q_{\parallel} \simeq \left(\frac{2\pi}{f}k \cos \varphi +  
\frac{2\pi}{h}l \sin \varphi\right) \vartheta_0,
\label{5.4c}
\end{equation} 
where $f$ and $h$ are the characteristic periods of the potential in the
plane transverse to the considered axis, $\varphi$ is the angle of the 
projection $\textbf{n}$ onto this plane with respect to one of the 
planes containing the selected axis, $k$ and $l$ are integers.

Let us consider the spectral distribution of pair production
in the extreme limit when the parameter 
$s=2\omega |q_{\parallel}|_{min}/m^2 \sim \omega \vartheta_0/m^2a_s \gg 1$.
In this case the maximum of distribution is attained at such values of 
$\vartheta_0$ where the standard CPP becomes inapplicable. Bearing in mind
that if $s \gg 1$ and $\vartheta_0 \sim V_0/m$ than  $\kappa_s \sim s \gg 1$,
we can conveniently  use the modified theory of CPP. Utilization of
the modified theory for these values of $s$ and $\varrho$
gives the exact position of maximum of the spectral distribution
(in region where $\tau \sim \tau_0$) and the value of the 
total probability within logarithmic accuracy $(\ln s \gg 1)$.

The transverse component of the vector $\textbf{q}$ in Eq.(\ref{5.4b})
can be selected in a such way that the spectral distribution given
by Eq.(\ref{5.4}) has a sharp maximum near the end of the spectrum
at $y_m(1-y_m)=(2+\varrho)/2s$ with relatively narrow (in terms
$1/s$) width $\Delta y \sim (1+\varrho/2)/s$
\begin{equation}
\left(\frac{dW_{\xi_+}}{dy} \right)_{max} =
\frac{\alpha \varrho |q_{\parallel}|_{min}}{4(2+\varrho)}
\left(1+\xi_+ + \frac{1-\xi_+}{\tau_0^2} \right),\quad
\tau_0=\frac{2s}{2+\varrho}.  
\label{6.4}
\end{equation} 
It is seen that the maximum of the spectral distribution with
the opposite helicity ($\xi_+=-1$) is suppressed as $1/\tau_0^2$.
At $\tau > \tau_0$ one have to take into account the next modes of
$q_{\parallel}$ (see Eq.(\ref{5.4b})). In this part of spectrum
the suppression of the probability with opposite helicity
is more strong, so the created positrons have nearly 
complete longitudinal polarization.

Bearing in mind that $\Delta y \sim (2+\varrho)/2s$ we find that
for $\xi_+=1$
\begin{equation}
\frac{dN_+}{dt} \sim \Delta y\left(\frac{dW_{\xi_+}}{dy} \right)_{max} \sim
\frac{\alpha \varrho m^2}{8\omega} \sim \frac{\varrho}{L_{rad}}
\frac{\varepsilon_e}{\omega},
\label{7.4}
\end{equation} 
where $L_{rad}$ is the radiation length in a corresponding 
amorphous medium, 
$\varepsilon_e=m(16 \pi Z^2 \alpha^2 n_a \lambda_c^3L_0)^{-1},
~L_0=\ln(183 Z^{-1/3}-f(Z\alpha))$, the function $f(Z\alpha))$
is defined in Eq.(\ref{4.3}), $N_+$ is the number of created positrons,
for tungsten one has $\varepsilon_e \simeq 2.5~$TeV. Thus, the above
analysis shows that the considered mechanism of creation of longitudinally
polarized positrons is especially effective because there is a gain both in the
monochromaticity and the total yield of polarized positrons with the
energy $\varepsilon_+ \simeq \omega$.

\section{Conclusion}
\setcounter{equation}{0}

It is shown above that at high energy in the process of
production of electron-positron pair with longitudinally polarized particles
by the circularly polarized photon in an oriented crystal the phenomenon
of helicity transfer takes place in the case when the final particle 
takes away nearly all energy of the photon.
This is true in the constant field limit $\vartheta_0 \ll V_0/m$
as well as in the coherent pair production region $\vartheta_0 > V_0/m$. 

In crossing channel: the radiation from 
longitudinally polarized high energy electrons in oriented crystals is circularly
polarized ($\xi^{(2)} \rightarrow 1$) near the end of spectrum \cite{BK4}
also in both regions.
This is once more the particular case of helicity transfer.

So, the oriented crystal is a very effective device for helicity transfer
from a photon to electron or positron and back from an electron to photon.
Near the end of spectrum this is nearly 100\% effect.

\vspace{0.5cm}

{\bf Acknowledgments}

The authors are indebted to the Russian Foundation for Basic
Research supported in part this research by Grant 
03-02-16154.

\newpage

\setcounter{equation}{0}
\Alph{equation}

\appendix

\section{Appendix}

{\large {\bf Integral polarization of positron in constant field, asymptotic}}
\vskip3mm
It is instructive that from the expression for $R_0(\lambda)$ (Eqs.(\ref{7.2}), (\ref{6.2})) 
one can find integral degree of positron longitudinal polarization 
in an external field characterized by parameter $\kappa$ for any $\omega$:
this is ratio of coefficient at $\xi_+$ and the rest part which enters 
in the probability for unpolarized particles
\begin{eqnarray}
\hspace{-8mm}&& \zeta = \frac{F_1}{F},\quad F_1
=\int_{0}^{1} \frac{dy}{y}\left[\left(\frac{y}{1-y}-1 \right)
K_{2/3}(\lambda) +\int_{\xi}^{\infty}K_{1/3}(z)dz \right],
\quad \xi=\frac{2}{3\kappa y(1-y)},
\nonumber \\
\hspace{-8mm}&& F=\int_{0}^{1} dy\left[\left(\frac{y}{1-y}+\frac{1-y}{y} \right)
K_{2/3}(\lambda) +\int_{\xi}^{\infty}K_{1/3}(z)dz \right], 
\label{a.1} 
\end{eqnarray} 
where $y=\varepsilon_+/\omega$. This result follows also 
from Eq.(3.70) \cite{BKS}. The integrand in the first term of $F_1$ is 
antisymmetric at substitution $y \leftrightarrow (1-y)$ and because of this
doesn't contributes into integral polarization.

Let us find the asymptotic values of $\zeta$. At $\kappa \ll 1$
it is convenient to write the function $F_1$ in the form
\begin{equation}
F_1=\int_{0}^{1} \frac{dy}{y}\int_{a/(4y(1-y))}^{\infty}K_{1/3}(z)dz
=\int_{1}^{\infty}\frac{dx}{\sqrt{x(x-1)}}\int_{ax}^{\infty}K_{1/3}(z)dz,
\label{a.2}
\end{equation}
where $a=8/3\kappa,~x=1/(4y(1-y))$. In the limit $a \gg 1$ one can
use the standard expansion of $K_{1/3}(z)$ for $z \gg 1$ and the
last integral in Eq.(\ref{a.2}) becomes
\begin{equation}
\int_{ax}^{\infty}K_{1/3}(z)dz \simeq \sqrt{\frac{\pi}{2}}\int_{ax}^{\infty}
\frac{e^{-z}}{\sqrt{z}}dz \simeq \sqrt{\frac{\pi}{2}}\frac{e^{-ax}}{\sqrt{ax}}.
\label{a.3}
\end{equation}
Taking the integral over $x$ we find
\begin{equation}
F_1 \simeq \frac{\pi}{\sqrt{2}a}e^{-a},
\label{a.4}
\end{equation}
while the function $F$ Eq.(\ref{a.1}) in this terms 
(see Eq.(3.58) in \cite{BKS}) is
\begin{equation}
F \simeq \frac{3\pi}{2\sqrt{2}a}e^{-a}.
\label{a.5}
\end{equation}
So, we find
\begin{equation}
\kappa \ll 1,\quad \zeta=\frac{2}{3}
\label{a.6}
\end{equation}
In the limit $\kappa \gg 1$ we present the integral $F_1$ as
\begin{eqnarray}
\hspace{-12mm}&& F_1=\int_{0}^{1} \frac{dy}{y}\int_{a/4y(1-y)}^{\infty}K_{1/3}(z)dz
=F_1^{(1)}+F_1^{(2)},\quad a \ll 1,\quad y_0 \ll 1,\quad \frac{a}{y_0} \ll 1,
\nonumber \\
\hspace{-12mm}&& F_1^{(1)}=\int_{y_0}^{1} \frac{dy}{y}
\int_{a/4y(1-y)}^{\infty}K_{1/3}(z)dz
\simeq \int_{y_0}^{1} \frac{dy}{y}\int_{0}^{\infty}K_{1/3}(z)dz
=\ln \left(\frac{1}{y_0}\right) \frac{\pi}{\sqrt{3}},
\nonumber \\
\hspace{-12mm}&& 
F_1^{(2)}=\int_{0}^{y_0}\frac{dy}{y}\int_{a/4y(1-y)}^{\infty}K_{1/3}(z)dz 
\simeq\int_{0}^{y_0}\frac{dy}{y}\int_{a/4y}^{\infty}K_{1/3}(z)dz=
\int_{a/4y_0}^{\infty}\frac{ds}{s}\int_{s}^{\infty}K_{1/3}(z)dz
\nonumber \\
\hspace{-12mm}&& 
\simeq -\ln\left(\frac{a}{4y_0}\right)\frac{\pi}{\sqrt{3}}+
\int_{0}^{\infty}\ln s K_{1/3}(s)ds
\nonumber \\
\hspace{-12mm}&&
=\frac{\pi}{\sqrt{3}}\left[-\ln \left(\frac{a}{4y_0}\right)+\frac{1}{2}
\left(\psi\left(\frac{1}{3}\right)+\psi\left(\frac{2}{3}\right)\right)
+\ln 2\right],
\label{a.7}
\end{eqnarray}
where $\psi(x)$ is the logarithmic derivative of the gamma function.
In calculation of $F_1^{(2)}$ we substitute the variable $s=a/4y$ and than
performed integration by parts.
Finally we have for $F_1$ in the limit $\kappa \gg 1$
\begin{equation}
F_1=\frac{\pi}{\sqrt{3}}\left(\ln \frac{\kappa}{\sqrt{3}}-C\right).
\label{a.8}
\end{equation} 
Using the two first terms of decomposition of total probability
of pair creation (see footnote at p.86 in \cite{BKS}) we obtain for
integral probability of polarized pair creation:
\begin{equation}
W^M(\xi_+)=\frac{\alpha m^2}{2\omega}\left[D\kappa^{2/3}-\frac{2}{3}
+\frac{\xi_+}{3}\left(\ln \frac{\kappa}{\sqrt{3}}-C\right)\right],
\label{a.9}
\end{equation}  
where 
\begin{equation}
D=\frac{5\Gamma(5/6)(2/3)^{1/3}}{14\Gamma(7/6)}=0.37961.
\label{a.10}
\end{equation} 
So we have for integral degree of longitudinal polarization
of created positrons
\begin{equation}
\zeta(\kappa) =\frac{\ln \kappa- 1.126}{1.139\kappa^{2/3}-2}.
\label{a.11}
\end{equation} 

\section{Appendix}
\setcounter{equation}{0}

{\large {\bf Coherent integral polarization term in a crystal 
at $\kappa_s \gg 1$}}
\vskip3mm

Here we consider situation when the value of parameter $\kappa(x)$
on the boundary of cell is small: $\kappa(x_0) \simeq 2\kappa_s/x_0^{3/2}
\ll 1$ and one can extend the integration interval over $x$ up to
infinity. The term $\propto \xi_+$ in the integral probability contains the
integral (see Appendix A, Eq.(\ref{a.2}))
\begin{equation}
F_{ax}=\int_{0}^{\infty}dx\int_{0}^{1}\frac{dy}{y}\int_{g}^{\infty}K_{1/3}(z)dz,\quad
a(x)=\frac{8}{3\kappa(x)}\simeq \frac{4(x+1)\sqrt{x}}{3\kappa_s},
\label{b.1}
\end{equation}
where $g=a(x)/(4y(1-y))$
The integration interval over $x$ we split into two parts
\[
1.~x \geq x_c,\quad 2.~x \leq x_c,
\]
where $\kappa_s^{3/2} \gg x_c \gg 1$. In the first interval 
 we have
\begin{eqnarray}
\hspace{-12mm}&&F_{ax}^{(1)}\simeq \int_{x_c}^{\infty}dx\int_{0}^{1} 
\frac{dy}{y}\left[ \int_{g_1(x)}^{\infty}K_{1/3}(z)dz
-\frac{\sqrt{x}}{3\kappa_s y(1-y)}K_{1/3}\left(g_1(x)\right) \right]
\nonumber \\
\hspace{-12mm}&&=\left[\int_{0}^{\infty}dx-\int_{0}^{x_c}dx \right]
\int_{0}^{1} \frac{dy}{y} \int_{g_1(x)}^{\infty}K_{1/3}(z)dz 
 -\frac{2}{3}\int_{0}^{1} \frac{dy}{y} 
\int_{g_1(x_c)}^{\infty}K_{1/3}(z)dz,
\nonumber \\
\hspace{-12mm}&& = F_{ax}^{(11)}+F_{ax}^{(12)}+F_{ax}^{(13)},
\label{b.2}
\end{eqnarray}
where $g_1(x)=x^{3/2}/(3\kappa_s y(1-y))$. In the first integral 
over $x:~F_{ax}^{(11)}=\int_{0}^{\infty}dx\ldots$ we reverse the
integration order, introduce the variable $s=g_1(x)$ and
integrate over $s$ by parts. We get
\begin{equation}
F_{ax}^{(11)}=(3\kappa_s)^{2/3}\int_{0}^{1}[y(1-y)]^{2/3}\frac{dy}{y}
\int_{0}^{\infty}s^{2/3}K_{1/3}(s)ds=\frac{3^{5/3}}{2^{4/3}}
\frac{\Gamma^{3}(2/3)}{\Gamma(1/3)}\kappa_s^{2/3}.
\label{b.3}
\end{equation}
In calculation of integrals over $y$ in the second interval over $x$, 
as well as in remaining integrals in Eq.(\ref{b.2}) we use
the results of previous Appendix. We have
\begin{eqnarray}
&&F_{ax}^{(2)}+F_{ax}^{(12)}+F_{ax}^{(13)} 
=-\frac{\pi}{\sqrt{3}}\left[\int_{0}^{x_c}
\left(\ln \frac{1+x}{x}\right)dx +\frac{2}{3}\left(
\ln \frac{2\kappa_s}{\sqrt{3x_c^3}} -C\right)  \right] 
\nonumber \\
&&\simeq
-\frac{2\pi}{3\sqrt{3}}
\left(\ln \frac{2\kappa_s}{\sqrt{3}}-C+\frac{3}{2} \right). 
\label{b.4}
\end{eqnarray}
Adding Eq.(\ref{b.3}) and Eq.(\ref{b.4}) we find
\begin{eqnarray}
\hspace{-15mm}&&F_{ax}
=c\kappa_s^{2/3}\left[1-
\frac{c_1}{\kappa_s^{2/3}}\left(\ln \kappa_s +c_2\right)\right],\quad 
c=\frac{3^{5/3}}{2^{4/3}}
\frac{\Gamma^{3}(2/3)}{\Gamma(1/3)}\simeq 2.295280..
\nonumber \\
\hspace{-15mm}&&
c_1=\frac{4\pi 2^{1/3}}{27\cdot 3^{1/6}}
\frac{\Gamma(1/3)}{\Gamma^{3}(2/3)} \simeq 0.526820..,~
c_2=\ln 2-\frac{\ln 3}{2}-C+\frac{3}{2} \simeq 1.066625..
\label{b.5}
\end{eqnarray}

\section{Appendix}
\setcounter{equation}{0}

{\large {\bf Incoherent integral polarization term in crystal 
at $\kappa_1 \gg 1$}}
\vskip3mm

At integration over $y$ the term containing $f_2(x,y)$ in Eq.(\ref{6.3})
vanishes due to symmetry of integrand at $y \leftrightarrow 1-y$
so the integral polarization term (containing $\xi_+$) is
\begin{equation}
F_{inc}=\int_{0}^{\infty}e^{-x/\eta_1}\frac{dx}{\eta_1}
\int_{0}^{1}f_1(x, y)dy,
\label{d.1}
\end{equation}
Let us consider the integral
\begin{equation}
F_1=\int_{0}^{1}f_1(x, y)dy=2\int_{0}^{1/2}f_1(z(x, y))dy,\quad
z=[\kappa y(1-y)]^{-2/3},
\label{d.2}
\end{equation}
where $f_1(z)$ is defined in Eq.(\ref{7.3}). Using the differential equation 
for Hardy function Eq.(\ref{c.6}) we can represent $f_1(z)$ in the form
\begin{equation}
f_1(z)=z^3\Upsilon''(z)-3z^2\Upsilon'(z).
\label{d.3}
\end{equation}
Let us split the integration interval in Eq.(\ref{d.2}) into two parts
\[
1.~0 \leq y \leq y_0, \quad 2.~y_0 \leq y \leq 1/2;\quad 
\kappa^{-1} \ll y_0 \ll 1.
\]
In the first interval $z \simeq (\kappa y)^{-2/3}$ and the 
corresponding integral is
\begin{eqnarray}
&&F_1^{(1)}=\frac{3}{\kappa}\int_{z_0}^{\infty}\left(z^3\Upsilon''(z)
-3z^2\Upsilon'(z) \right) \frac{dz}{z^{5/2}}
\nonumber \\
&&=\frac{3}{\kappa}\left[\int_{0}^{\infty}dz-\int_{0}^{z_0}dz \right]
\left(\sqrt{z}\Upsilon''(z)-\frac{3}{\sqrt{z}}\Upsilon'(z) \right),
\quad z_0=(\kappa y_0)^{-2/3} \ll 1.  
\label{d.4}
\end{eqnarray}
Now we turn to the first integral
\begin{equation}
\int_{0}^{\infty}\left(\sqrt{z}\Upsilon''(z)-
\frac{3}{\sqrt{z}} \Upsilon'(z)\right)dz=-\frac{7}{2}
\int_{0}^{\infty}\Upsilon'(z)\frac{dz}{\sqrt{z}}. 
\label{d.5}
\end{equation}
Using the representation Eq.(\ref{8.3}) we get
\begin{eqnarray}
&&\int_{0}^{\infty}\Upsilon'(z)\frac{dz}{\sqrt{z}}=
\int_{0}^{\infty}\frac{dz}{\sqrt{z}}\int_{0}^{\infty}
\tau \cos\left(z\tau+\frac{\tau^3}{3} \right)d\tau=
2 \int_{0}^{\infty}dx\int_{0}^{\infty}
\tau \cos\left(x^2\tau+\frac{\tau^3}{3} \right)d\tau
\nonumber \\
&&=2\int_{0}^{\infty}dx\int_{0}^{\infty}\sqrt{\tau} \cos\left(x^2+\frac{\tau^3}{3} \right)d\tau
=\frac{4}{\sqrt{3}}\int_{0}^{\infty}dy\int_{0}^{\infty}dx
\cos(x^2+y^2)
\nonumber \\
&&=\frac{2\pi}{\sqrt{3}}\int_{0}^{\infty}r\cos(r^2)dr
=\frac{\pi}{\sqrt{3}}\int_{0}^{\infty}\cos s ds=0.
\label{d.6}
\end{eqnarray}
The second integral in Eq.(\ref{d.4}) is
\begin{eqnarray}
&&\int_{0}^{z_0}\left(\sqrt{z}\Upsilon''(z)-
\frac{3}{\sqrt{z}} \Upsilon'(z)\right)dz \simeq
\int_{0}^{z_0}\left(\sqrt{z}\Upsilon''(0)-
\frac{3}{\sqrt{z}} \Upsilon'(0)\right)dz
\nonumber \\
&&=\frac{2}{3}z_0^{3/2}-6z_0^{1/2}\Upsilon'(0).
\label{d.7}
\end{eqnarray}
In the second interval $y_0 \leq y \leq 1/2$ the 
variable $z \ll 1$ than the integral in Eq.(\ref{d.2}) is
\begin{equation}
F_1^{(2)} \simeq -2\int_{y_0}^{1/2}\left[\frac{1}{\kappa^2 y^2(1-y)^2}
+\frac{3\Upsilon'(0)}{\kappa^{4/3}y^{4/3}(1-y)^{4/3}} \right]dy. 
\label{d.8}
\end{equation}
In the first term ($\propto \kappa^{-2}$) we retain only the main term 
$\propto 1/y_0$ and in the second term we carry out the integration 
by parts. We find
\begin{equation}
F_1^{(2)} \simeq -\frac{2}{\kappa^2 y_0} -
\frac{18\Upsilon'(0)}{\kappa^{4/3}y_0^{1/3}}+
\frac{18\cdot 2^{5/3}\Upsilon'(0)}{\kappa^{4/3}}-
\frac{24\Upsilon'(0)}{\kappa^{4/3}}\int_{0}^{1/2}
\frac{dy}{y^{1/3}(1-y)^{7/3}}.
\label{d.9}
\end{equation}
Substituting Eq.(\ref{d.5})-Eq.(\ref{d.7}) into Eq.(\ref{d.4})
and combining the result with Eq.(\ref{d.9}) we obtain
\begin{eqnarray}
&&F_1=F_1^{(1)}+F_1^{(2)} \simeq \frac{24\Upsilon'(0)}{\kappa^{4/3}}
\left(3\cdot2^{-1/3}-b_1 \right), 
\nonumber \\
&&b_1=\int_{0}^{1/2}
\frac{dy}{y^{1/3}(1-y)^{7/3}}= 1.86775..,\quad F_1\simeq \frac{5.7838}{\kappa^{4/3}}.
\label{d.10}
\end{eqnarray}
Integrating in Eq.(\ref{d.1}) over $s=x/\eta_1~(\kappa(s)
\simeq \kappa_1 \sqrt{2s}/(1+s))$ and using the numerical
value of the integral
\begin{equation}
2^{-2/3}\int_{0}^{\infty}e^{-s}s^{-2/3}(1+s)^{4/3}=2.561..,
\label{d.11}
\end{equation}
we find 
\begin{equation}
\sigma_{\xi_+} \simeq 3.95 \frac{Z^2\alpha^3L(\kappa_1)}{m^2\kappa_1^{4/3}}.
\label{d.12}
\end{equation}

\section{Appendix}
\setcounter{equation}{0}

{\large {\bf The Hardy function}}
\vskip3mm
The Hardy function
\begin{equation}
\Upsilon(z)=\int_{0}^{\infty} \sin\left(z\tau+
\frac{\tau^3}{3} \right) d\tau
\label{c.1}
\end{equation}
is encountered in the theory of electromagnetic processes in
an external field.

At $z \ll 1$ the decomposition of $\Upsilon(z)$ is
\begin{equation}
\Upsilon(z)=\frac{1}{3^{2/3}}\sum_{0}^{\infty}\frac{(-3^{1/3}z)^k}{k!}
\Gamma\left(\frac{k+1}{3} \right)\cos \left(\frac{k+1}{3} \pi\right)
=\frac{\Gamma(1/3)}{2\cdot 3^{2/3}}+\frac{\Gamma(2/3)}{2\cdot 3^{1/3}}z-
\frac{z^2}{2}+...,
\label{c.2}
\end{equation}
and $\Upsilon(0)=0.643950..$, $\Upsilon'(0)=0.469447..$.

At $z \gg 1$ there is asymptotic series of $\Upsilon(z)$ over
$1/z^3$:
\begin{equation}
\Upsilon(z)=\frac{1}{z}\sum_{0}^{\infty}\frac{(3k)!}{k!}
\frac{1}{(3z^3)^k}=\frac{1}{z}\left(1+\frac{2}{z^3}+
\frac{40}{z^6}+\frac{2240}{z^9}+... \right). 
\label{c.3}
\end{equation}

For calculation it is convenient to use the following 
representation of the Hardy function and its derivative
\begin{eqnarray}
&& \Upsilon(z)=\int_0^{\infty}\sin
\left(\frac{\sqrt{3}}{2}z\tau+\frac{\pi}{6}\right)
\exp\left(-\frac{z\tau}{2}-\frac{\tau^3}{3}\right) d\tau, 
\nonumber \\
&& \Upsilon'(z)=\int_0^{\infty}\cos
\left(\frac{\sqrt{3}}{2}z\tau+\frac{\pi}{6}\right)
\exp\left(-\frac{z\tau}{2}-\frac{\tau^3}{3}\right)\tau d\tau. 
\label{c.4}
\end{eqnarray}
It can be obtained using the expression
\begin{equation}
\Upsilon(z)={\rm Im}\int_{0}^{\infty} \exp\left( i\left(z\tau+
\frac{\tau^3}{3} \right)\right)  d\tau
\label{c.5}
\end{equation}
after a turn of integration line by the angle $\pi/6$.

The Hardy function satisfies the equation
\begin{equation}
\Upsilon''(z)-z\Upsilon(z)=-1.
\label{c.6}
\end{equation}

\newpage

{\bf Figure captions}

\vspace{8mm}
\begin{itemize}

\item {\bf Fig.1} The spectral probability of pair creation $dw_{\xi_+}^F/dy$,
the curves 1 and 2 are for energy $\omega$=22~GeV,
the curves 3 and 4 are for energy $\omega$=100~GeV,
the curves 5 and 6 are for energy $\omega$=250~GeV.
The curves 1, 3 and 5 are for $\xi=1$, and
the curves 2, 4 and 6 are for $\xi=-1$. 

\item {\bf Fig.2} The integral probability of pair creation by a photon in
tungsten, axis $<111>$, $T=100~$ K as function of photon energy.
The curve 1 is incoherent contribution, the curve 2 is coherent contribution  
the curve 3 is their sum giving the total probability. 

\item {\bf Fig.3} The positron polarization vs relative positron energy 
$\varepsilon/\omega$ in tungsten, axis $<111>, T=100~K$. 
The curve 1 for $\omega=12~$GeV,
the curve 2 for $\omega=22~$GeV, the curve 3 for $\omega=100~$GeV.
Both coherent and incoherent contributions are taken into account.

\item {\bf Fig.4}. The integral positron polarization $\xi_+$ as a function of 
photon energy $\omega$.

\end{itemize}

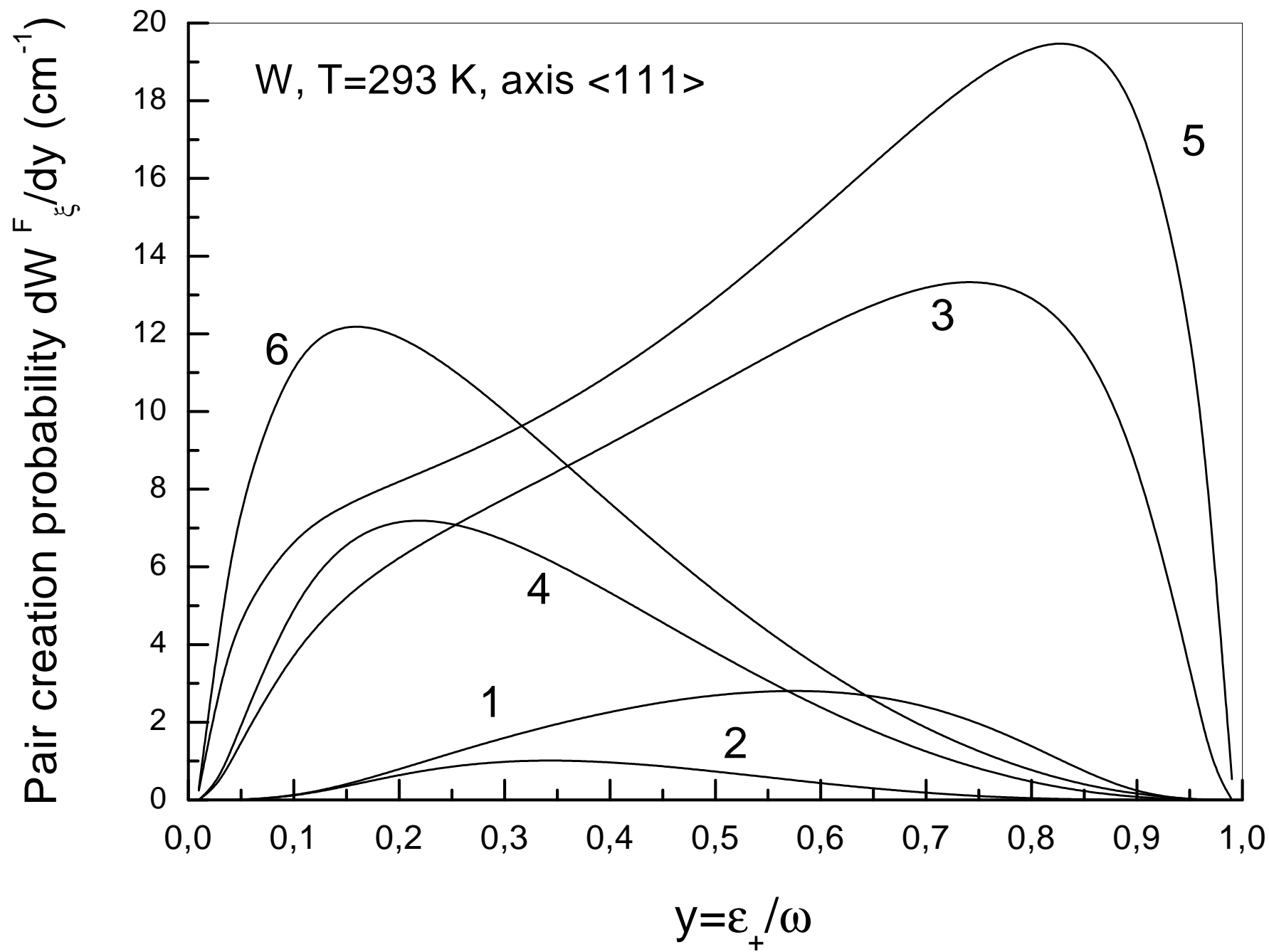
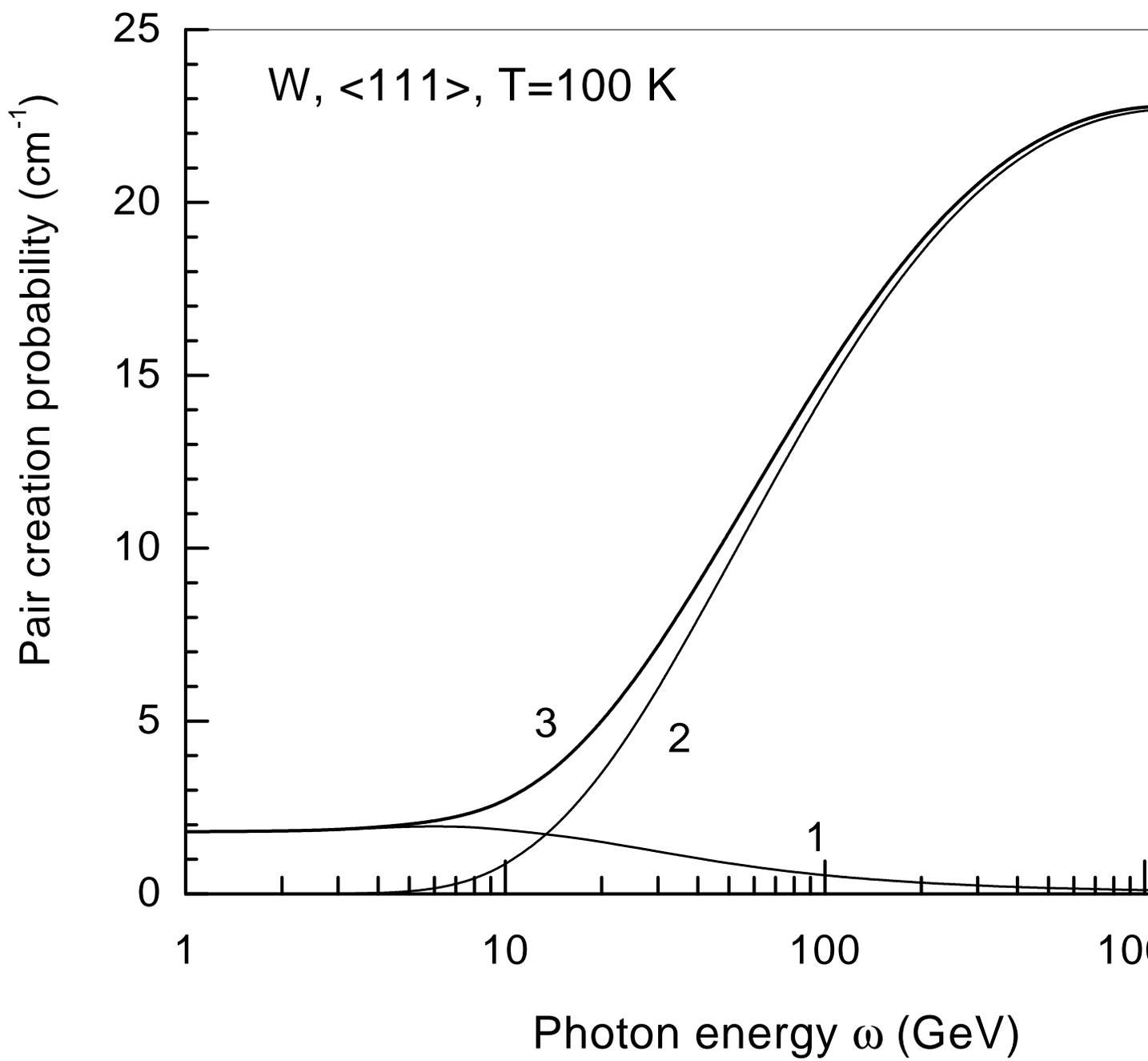
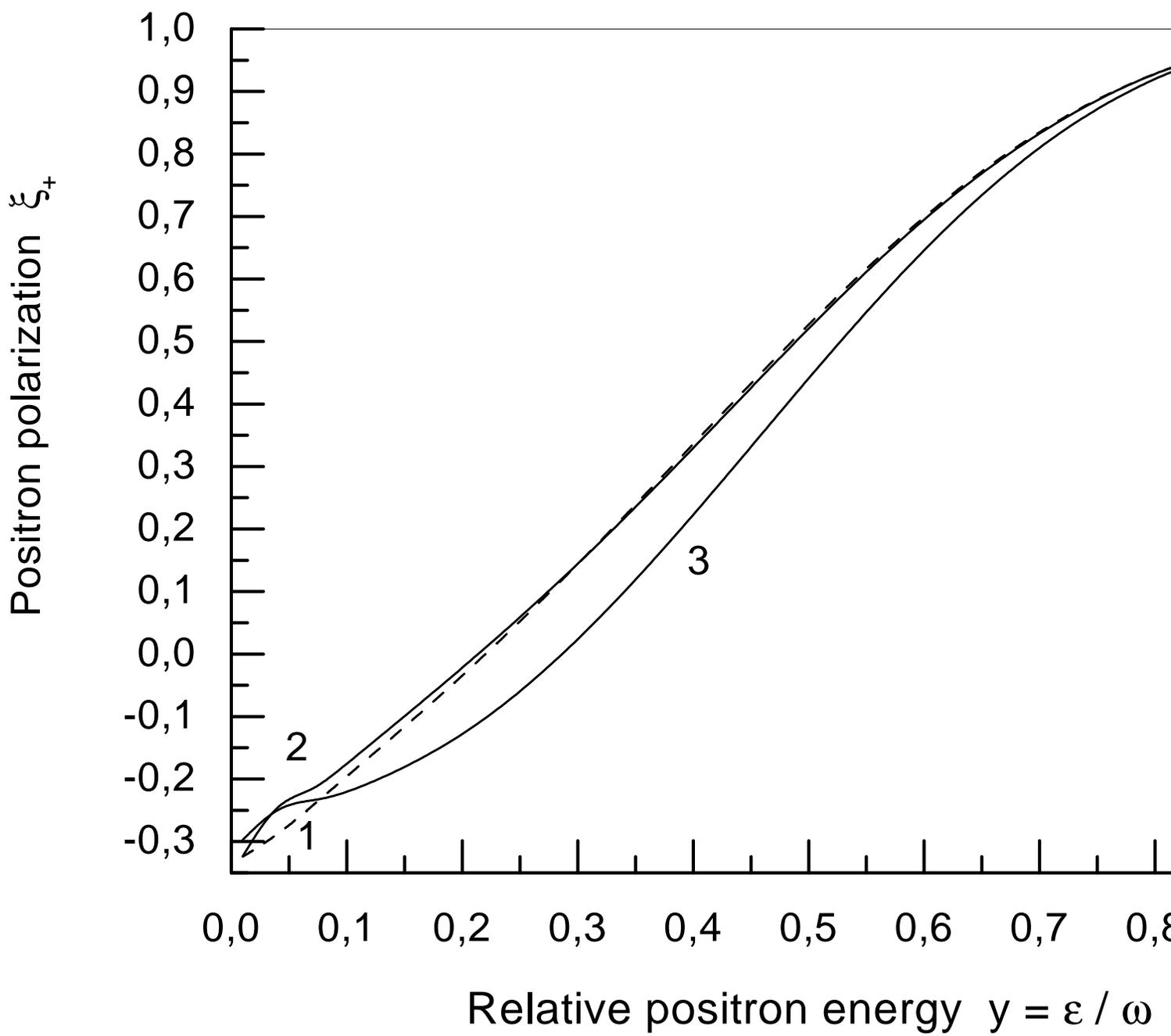
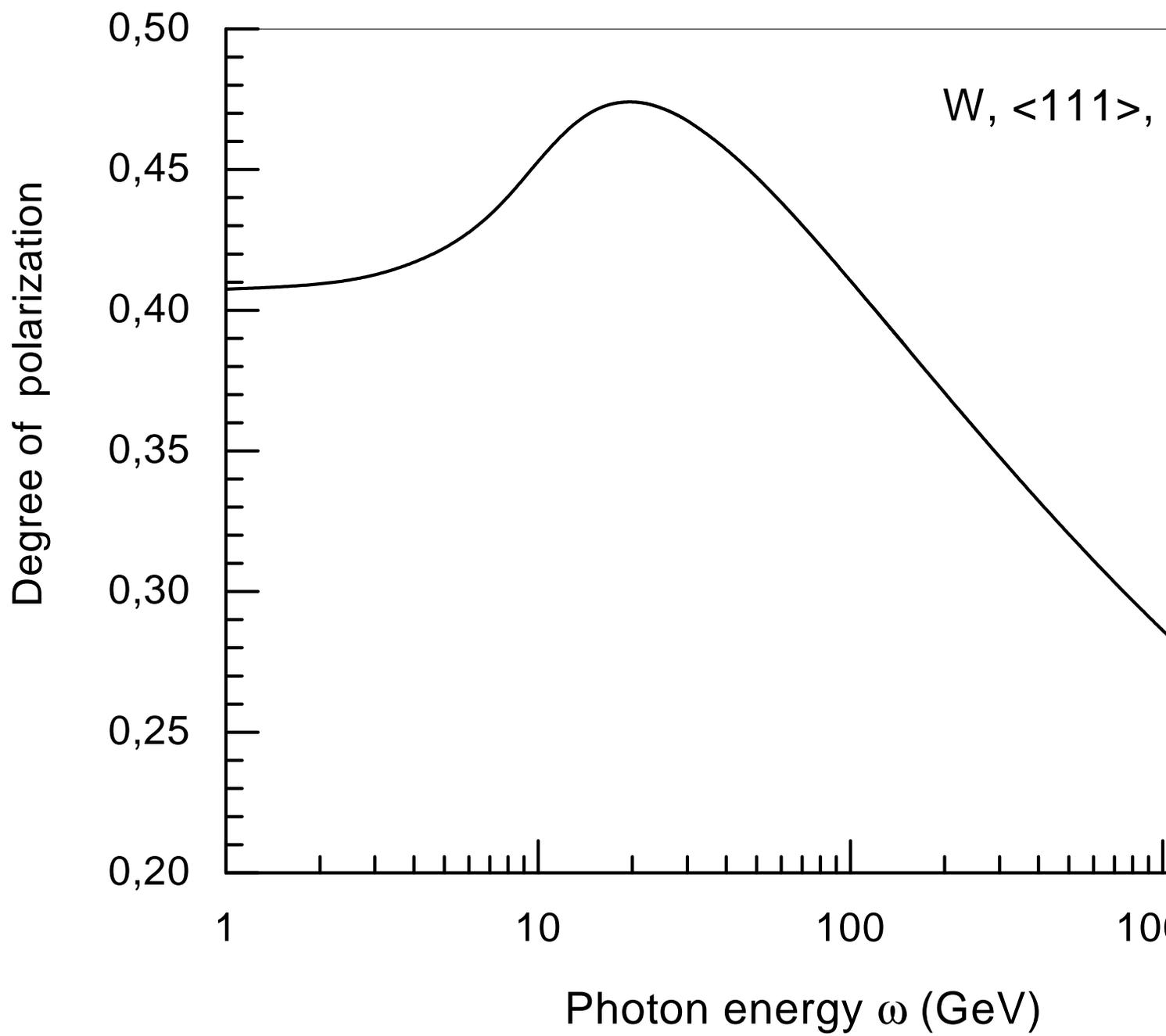
\end{document}